\documentclass[twocolumn,accepted=2019-03-30]{quantumarticle}
\usepackage{amsmath}
\usepackage[numbers]{natbib}
\usepackage{amssymb}
\usepackage{amsthm}
\usepackage{amsfonts}
\usepackage[pdfpagelabels,pdftex,bookmarks,breaklinks]{hyperref}
\usepackage[all]{hypcap}
\usepackage[section]{placeins}
\hypersetup{colorlinks,citecolor=blue,urlcolor=blue,linkcolor=blue}

\newcommand{\eq}[1]{Eq.~\hyperref[eq:#1]{(\ref*{eq:#1})}}
\renewcommand{\sec}[1]{\hyperref[sec:#1]{Section~\ref*{sec:#1}}}
\DeclareRobustCommand{\app}[1]{\hyperref[app:#1]{Appendix~\ref*{app:#1}}}
\newcommand{\fig}[1]{\hyperref[fig:#1]{Figure~\ref*{fig:#1}}}

\newcommand{\factory}[3]{$#1 \genfrac{}{}{0pt}{}{#2}{\rightarrow} {#3}$ factory}

\begin{document}

\title{\texorpdfstring{
Efficient magic state factories with a catalyzed $|CCZ\rangle \rightarrow 2|T\rangle$ transformation
}{
Efficient magic state factories with a catalyzed CCZ to 2T transformation
}}

\date{\today}
\author{Craig Gidney}
\email{craiggidney@google.com}
\affiliation{Google Inc., Santa Barbara, California 93117, USA}
\author{Austin G. Fowler}
\affiliation{Google Inc., Santa Barbara, California 93117, USA}

\begin{abstract}
We present magic state factory constructions for producing $|CCZ\rangle$ states and $|T\rangle$ states.
For the $|CCZ\rangle$ factory we apply the surface code lattice surgery construction techniques described in \cite{fowler2018} to the fault-tolerant Toffoli \cite{jones2013, eastin2013distilling}.
The resulting factory has a footprint of $12d \times 6d$ (where $d$ is the code distance) and produces one $|CCZ\rangle$ every $5.5d$ surface code cycles.
Our $|T\rangle$ state factory uses the $|CCZ\rangle$ factory's output and a catalyst $|T\rangle$ state to exactly transform one $|CCZ\rangle$ state into two $|T\rangle$ states.
It has a footprint 25\% smaller than the factory in \cite{fowler2018} but outputs $|T\rangle$ states twice as quickly.
We show how to generalize the catalyzed transformation to arbitrary phase angles, and note that the case $\theta=22.5^\circ$ produces a particularly efficient circuit for producing $|\sqrt{T}\rangle$ states.
Compared to using the $12d \times 8d \times 6.5d$ $|T\rangle$ factory of \cite{fowler2018}, our $|CCZ\rangle$ factory can quintuple the speed of algorithms that are dominated by the cost of applying Toffoli gates, including Shor's algorithm \cite{shor1994} and the chemistry algorithm of Babbush et al. \cite{babbush2018}.
Assuming a physical gate error rate of $10^{-3}$, our CCZ factory can produce $\sim 10^{10}$ states on average before an error occurs.
This is sufficient for classically intractable instantiations of the chemistry algorithm, but for more demanding algorithms such as Shor's algorithm the mean number of states until failure can be increased to $\sim 10^{12}$ by increasing the factory footprint $\sim 20\%$.
\end{abstract}

\maketitle

\section{Introduction}
\label{sec:introduction}

In fault-tolerant quantum computation based on the surface code (a likely component of future error corrected quantum computers due to the surface code's comparatively high threshold and planar connectivity requirements \cite{Brav98,Denn02,Raus07,Raus07d,Fowl12f}), the cost of a quantum algorithm is well approximated by the number of non-Clifford operations.
This is due to the fact that non-Clifford operations are performed via magic state distillation \cite{bravyi2005}, and the cost of state distillation is large.
For example, the spacetime volume (qubit-seconds) of the T state factory from \cite{fowler2018} is two orders of magnitude larger than the volume of a CNOT operation between adjacent qubits \cite{horsman2012}.
The non-Clifford gate count will likely be particularly significant for the earliest error corrected quantum computers, which will not have enough space to distill magic states in parallel.

\begin{figure*}
  \label{fig:overview-dataflow}
  \resizebox{\textwidth}{!}{
    \includegraphics{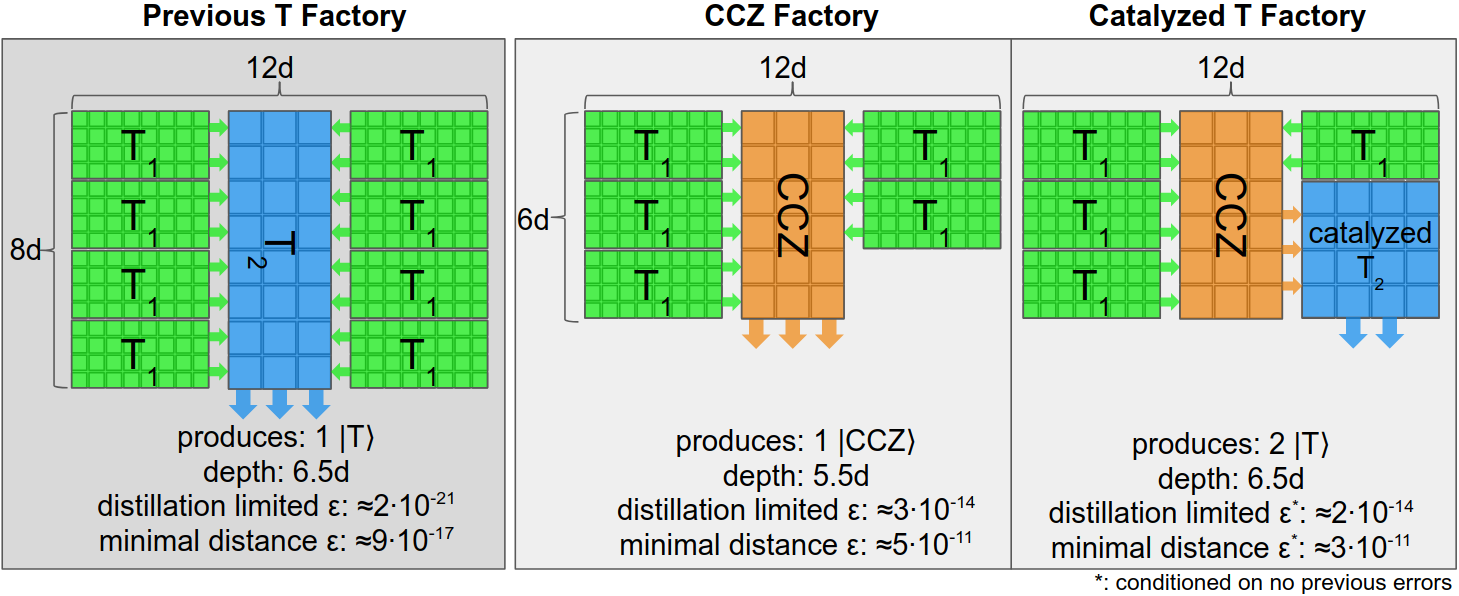}
  }
  \caption{
    Overview of the spatial layout and data flow of the \factory{15|T\rangle}{35 \epsilon^3}{|T\rangle} construction from \cite{fowler2018} (left), our \factory{8|T\rangle}{28\epsilon^2}{|CCZ\rangle} construction (middle), and our $|T\rangle$-catalyzed \factory{8|T\rangle}{28\epsilon^2}{2|T\rangle} (right).
    The level 1 $T$ factories (green) are effectively the same as in \cite{fowler2018}, and are performed at half code distance to balance the contributions from distillation error and code error.
    The distillation limited error rates assume a physical gate error rate of $10^{-3}$, that the injection technique of \cite{li2015} can create level 0 $|T\rangle$ states with approximately that level of error, and that the code distance is large enough for the dominant source of error in the outputs to be distillation error.
    The minimal distance error rates include the effects of topological errors in the surface code itself, with a code distance of 7 for level 0 $|T\rangle$ state injection, code distance 15 for level 1 factories, and code distance 31 for everything else.
    The factory from \cite{fowler2018} has significantly better error suppression, but the amount of suppression is overkill unless one wants to run century-long computations without a single error.
    Our factories have smaller footprints, faster output, and an amount of suppression sufficient to run proposed algorithms beyond the classically simulable regime (e.g. \cite{babbush2018}).
    The error rates of the catalyzed T factory have asterisks because its errors are correlated: if one error occurs it can poison the catalyst state and cause many more errors.
    This means that this factory should be used in contexts where a single error is already considered a complete failure (e.g. at the level of an entire algorithm, not as an input to further distillations).
  }
\end{figure*}

\begin{figure*}
  \label{fig:overview-3d}
  \resizebox{\textwidth}{!}{
    \includegraphics{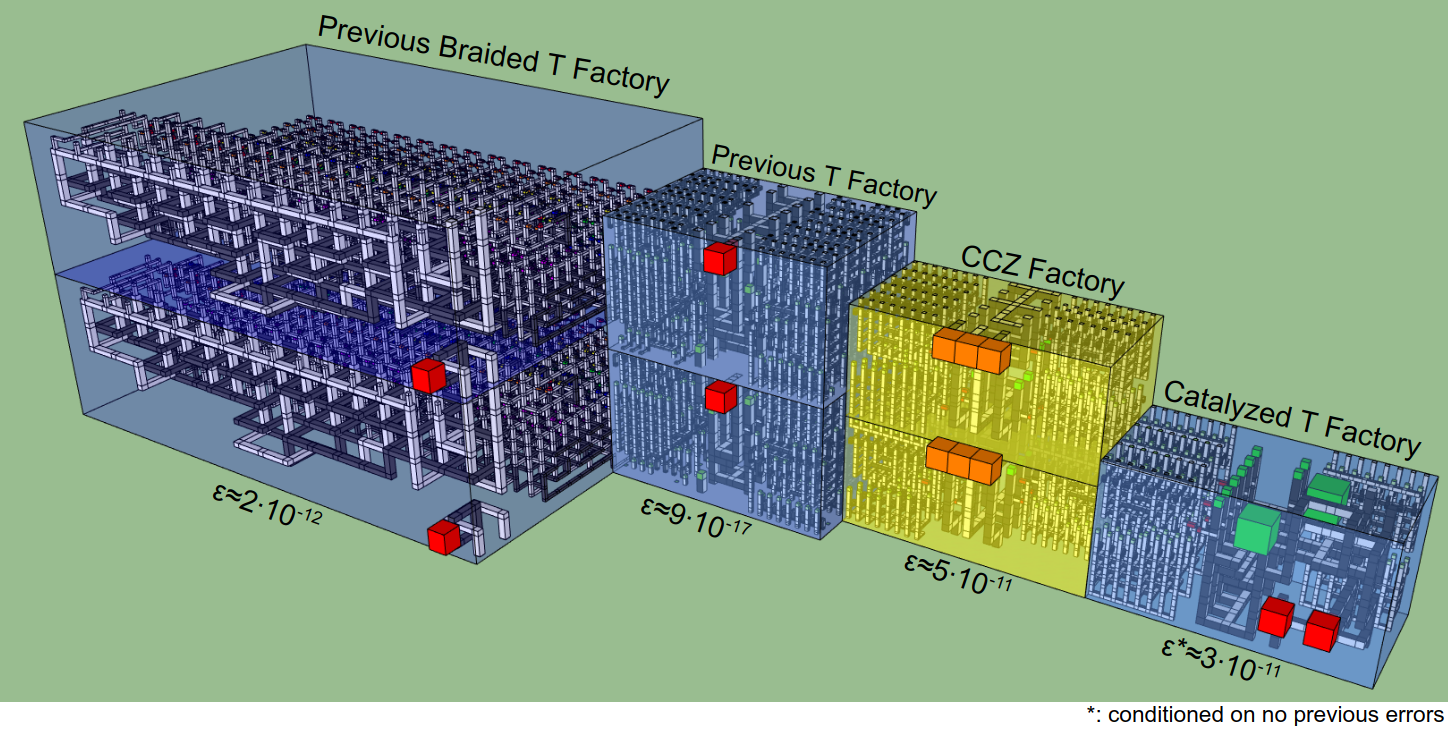}
  }
  \caption{
    Size comparison of various factories producing two magic states, including output error rates.
    Error rates are computed assuming a physical gate error rate of $10^{-3}$, and include topological errors from the surface code itself.
    Includes the \factory{15|T\rangle}{35 \epsilon^3}{|T\rangle} construction using braids \cite{fowler2012bridge} (left) and lattice surgery \cite{fowler2018} (middle left), as well as our \factory{8|T\rangle}{28\epsilon^2}{|CCZ\rangle} construction (middle right) and our $|T\rangle$-catalyzed \factory{8|T\rangle}{28\epsilon^2}{2|T\rangle} (right).
    $|T\rangle$ output events are indicated with red cubes.
    $|CCZ\rangle$ output events are indicated with a triplet of orange cubes.
    The braided factory has been scaled to account for the fact that it uses the unrotated surface code instead of the rotated surface code \cite{horsman2012}.
    The braided T factory's error rate is significantly higher because it uses an older injection technique, resulting in the level 0 T gates having an error rate of $10^{-2}$ instead of $2 \cdot 10^{-3}$.
    The error rate of the catalyzed T factory has an asterisk because its errors are correlated: if one error occurs it can poison the catalyst state and cause many more errors.
    This means that this factory should be used in contexts where a single error is already considered a complete failure (e.g. at the level of an entire algorithm, not as an input to further distillations).
  }
\end{figure*}

\begin{figure*}
  \label{fig:spreadsheet}
  \resizebox{\textwidth}{!}{
    \includegraphics{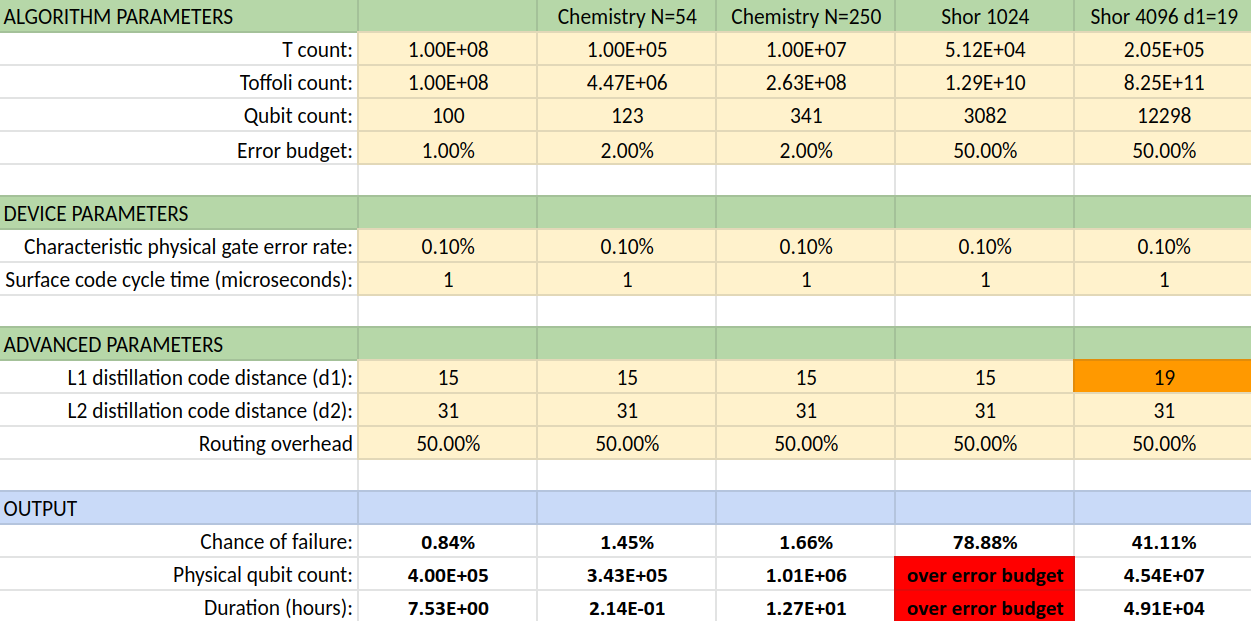}
  }
  \caption{
    Screenshot of the resource estimation spreadsheet include in the supplementary materials of this paper (file name ``calculator-CCZ-2T-resources.ods"), with various interesting cases pre-entered.
    Assuming a physical gate error rate of $10^{-3}$, and minimal code distances, the $|CCZ\rangle$ factory is unlikely to fail when producing on the order of $10^{10}$ states.
    This is sufficient to run classically intractable chemistry algorithms \cite{babbush2018}, but not quite sufficient to factor a 1024 bit number with a 50\% success rate (assuming that factoring an $n$ bit number requires $12 n^3$ Toffoli gates and $3n$ space \cite{zalka1998fast}).
    However, if the physical gate error rate is improved slightly or (more plausibly) the factory is made slightly larger by increasing the level 1 code distance from 15 to 19, then the number of states that can be produced increases to be on the order of $10^{12}$.
    This allows 4096 bit numbers to be factored (though we do not recommend using a single factory for this task, since it would take 5 years to produce the necessary magic states assuming a surface code cycle time of 1 microsecond).
  }
\end{figure*}

\begin{figure*}
  \label{fig:diagram-style-3d}
  \resizebox{\textwidth}{!}{
    \includegraphics{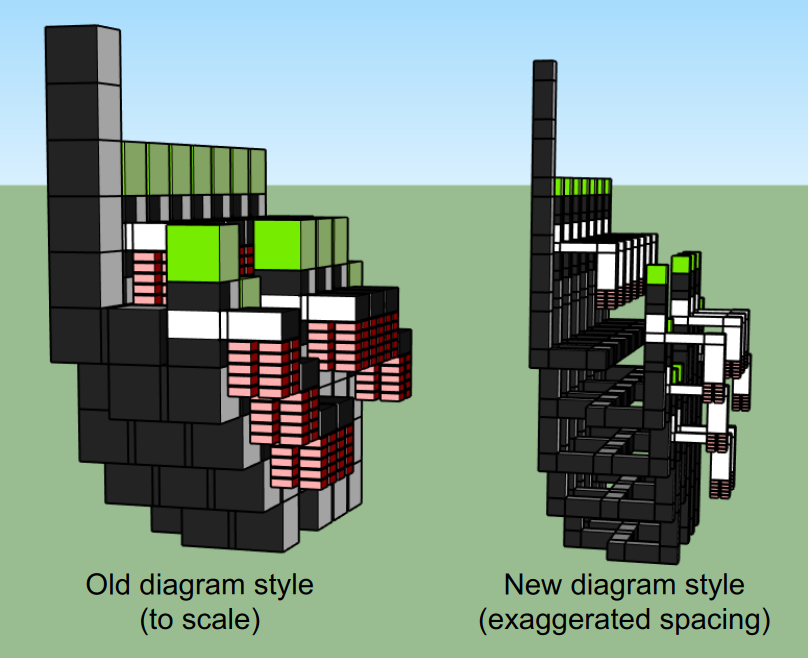}
  }
  \caption{
    Comparison of the to-scale diagram style from \cite{fowler2018} with the exaggerated-spacing diagram style used by this paper.
    The to-scale diagram style emphasizes how things fit together and is ideal when reasoning geometrically.
    The exaggerated-spacing diagram style emphasize how things connect together and is ideal when reasoning topologically.
    An even more abstract diagram style for lattice surgery is the ZX calculus \cite{de2017}.
    In \fig{ccz-graph} we show how translate a topological diagram into a ZX calculus graph.
  }
\end{figure*}

Over the past decade, thanks to techniques such as block codes \cite{bravyi2012, fowler2013}, bridge compression \cite{fowler2012bridge}, and many others \cite{horsman2012, campbell2017, campbell2018, litinski2018}, the cost of magic state distillation has steadily decreased.
This paper adds catalyzed phasing to the pile of known techniques, continuing the tradition of gradually chipping away at the convenient approximation that magic states are the dominant cost in error-corrected quantum computation.

Note that, in this paper, we focus on optimizing the cost of distillation in the {\em single-factory regime}.
For example, we do not investigate whether there are block code factories that can use catalyzation.
We focus on the single-factory regime because we are interested in estimating the minimum number of physical qubits needed to run classically intractable instances of various quantum algorithms at a reasonable rate, and the single-factory regime is the relevant one for these kinds of estimates.
In \fig{overview-dataflow} and \fig{overview-3d}, we give a high-level view of this paper's improvements in footprint and spacetime volume, over previous factories in the single-factory regime.

The paper is organized as follows.
\sec{introduction} provides an overview of the paper, and explains the various notation and diagram conventions we will be using.
In \sec{ccz}, we explain how to construct an efficient $|CCZ\rangle$ factory by applying the techniques of \cite{fowler2018} to the construction of \cite{jones2013, eastin2013distilling}.
In \sec{catalysis}, we construct a circuit which can transform a $|CCZ\rangle$ state into two $|T\rangle$ states if a catalyst $|T\rangle$ state is present.
In \sec{generalize}, we show that this catalyzed circuit generalizes to other phase angles and note that this generalized circuit can produce two $|\sqrt{T}\rangle$ states using only five $|T\rangle$ states.
In \sec{full}, we combine constructions from the previous sections into an efficient $|T\rangle$ factory.
Finally, in \sec{conclusion}, we discuss applications of our constructions, summarize our contributions, and point towards future work.

In this paper we will refer to factories using the notation ``\factory{|\text{In}\rangle}{f(\epsilon)}{|\text{Out}\rangle}".
The left hand side is the state input into the factory, the right hand side is the state output from the factory, and the function above the arrow indicates the amount of error suppression up to leading terms (i.e. the $f(\epsilon)$ above the arrow is shorthand for the true suppression $f(\epsilon) + O(\epsilon f(\epsilon))$).
For example, we will refer to the $|T\rangle$ state distillation based on the 15-qubit Reed-Muller code \cite{bravyi2005} as the \factory{15|T\rangle}{35 \epsilon^3}{|T\rangle}.

We use three main types of diagram in this paper: circuit diagrams, time slice diagrams, and 3D topological diagrams.
The circuit diagrams demonstrate the functionality that the 3D topological diagrams are supposed to be implementing, and the time slice diagrams are a sequence of slices through the 3D topological diagrams, showing boundary information and which patches are being merged or split.
We often provide multiple diagrams of the same construction, with common labelling between the diagrams.
For example, \fig{ccz-circuit}, \fig{ccz-slices}, and \fig{ccz-3d} are all diagrams of our CCZ factory.
Discerning readers can use the labels common to all three diagrams to verify that they agree with each other.
In particular, those three diagrams all have three output qubits labelled 1 through 3, eight ancillae qubits labelled $a$ through $h$, and four ``stabilizer qubits" labelled by the stabilizer measurement they correspond to.

To make it possible to see the internal topological structure of the 3D topological diagrams, we have chosen to significantly exaggerate the amount of space between events.
We draw operations as if they had linear $O(d)$ separation (where $d$ is the code distance), but on actual hardware the operations have a constant $O(1)$ separation.
This exaggeration of separation does not change the topology, so interpreting the figures as if they were to scale will still produce the correct computation.
But it is important to account for the distortion when computing the footprint or depth of the computation.
\fig{diagram-style-3d} shows a comparison between the old to-scale diagram style and our new exaggerated spacing style diagrams.

We will sometimes refer to multi-qubit stabilizers using a concatenated-subscript notation such as $Z_{123}$.
Each subscript refers to a separate qubit, i.e. $Z_{123} = Z_1 Z_2 Z_3$.

We will often refer to $|T\rangle$ states as having a particular ``level", e.g. ``a level 1 $|T\rangle$ state" or equivalently ``a $|T_1\rangle$ state".
The level refers to the number of distillation steps used to produce the state.
We will also refer to factories by the level of their output.
For example, our starting point is level 0 $|T\rangle$ states produced using the post-selected state injection of Li \cite{li2015}.
These $|T_0\rangle$ states are then distilled by the level 1 T factory from \cite{fowler2018} into $|T_1\rangle$ states, which we can then feed into our $|CCZ\rangle$ factory.

Lastly, we wish to point out the useful supplementary materials included with this paper.
First, because it is significantly easier to understand 3D diagrams when one is able to move the camera, the supplementary materials include SketchUp files storing the models shown in the 3D topological diagrams.
Second, the supplementary materials include a spreadsheet (file name ``calculator-CCZ-2T-resources.ods") that can compute the overhead of computations that use our factories.
Interested readers can estimate the running time and number of physical qubits required by their algorithms by entering into the spreadsheet the number of T and Toffoli gates performed by the algorithm, how many qubits the algorithm uses, and an error budget.
\fig{spreadsheet} shows a screenshot of the spreadsheet.

\section{\texorpdfstring{
    Lattice surgery construction of the \factory{8|T\rangle}{28\epsilon^2}{|CCZ\rangle}
}{
    Lattice surgery construction of the 8T to CCZ  factory
}}
\label{sec:ccz}

\begin{figure*}
  \label{fig:ccz-circuit}
  \resizebox{\textwidth}{!}{
    \includegraphics{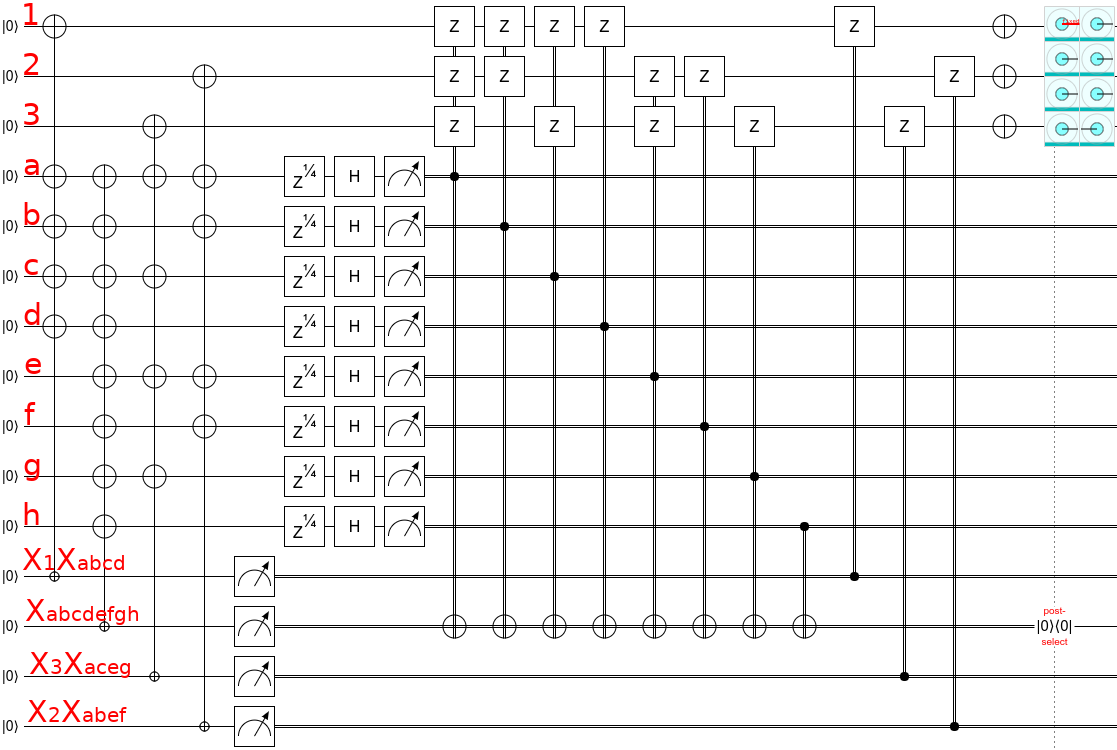}
  }
  \caption{
    Quantum circuit for the \factory{8|T\rangle}{28\epsilon^2}{|CCZ\rangle}.
    A rewrite of figure 3 from \cite{jones2013}.
    The box with blue circles in the top right is a state display from the online simulator Quirk, with each circle representing an amplitude (the radius of the colored circle indicates the amplitude's magnitude, and the angle of the line rooted at the center of the circle indicates the phase).
    The state display is showing that the output state is a $|CCZ\rangle$ state.
    The small circled pluses in the circuit are X-axis controls (equivalent to a normal control surrounded by Hadamard gates); whenever one of these controls directly precedes a measurement the measurement corresponds to a Pauli product measurement.
    The post-selection operation represents the classical control software determining if the an error was detected; if it fails the output must be discarded.
    Pauli operations and classically-controlled Pauli operations appear here, but not in \fig{ccz-3d}, because they are performed entirely within classical control software.
    The circuit can be opened in Quirk by \href{http://algassert.com/quirk\#circuit=\%7B\%22cols\%22\%3A\%5B\%5B\%22X\%22\%2C1\%2C1\%2C\%22X\%22\%2C\%22X\%22\%2C\%22X\%22\%2C\%22X\%22\%2C1\%2C1\%2C1\%2C1\%2C\%22\%E2\%8A\%96\%22\%5D\%2C\%5B1\%2C1\%2C1\%2C\%22X\%22\%2C\%22X\%22\%2C\%22X\%22\%2C\%22X\%22\%2C\%22X\%22\%2C\%22X\%22\%2C\%22X\%22\%2C\%22X\%22\%2C1\%2C\%22\%E2\%8A\%96\%22\%5D\%2C\%5B1\%2C1\%2C\%22X\%22\%2C\%22X\%22\%2C1\%2C\%22X\%22\%2C1\%2C\%22X\%22\%2C1\%2C\%22X\%22\%2C1\%2C1\%2C1\%2C\%22\%E2\%8A\%96\%22\%5D\%2C\%5B1\%2C\%22X\%22\%2C1\%2C\%22X\%22\%2C\%22X\%22\%2C1\%2C1\%2C\%22X\%22\%2C\%22X\%22\%2C1\%2C1\%2C1\%2C1\%2C1\%2C\%22\%E2\%8A\%96\%22\%5D\%2C\%5B1\%2C1\%2C1\%2C1\%2C1\%2C1\%2C1\%2C1\%2C1\%2C1\%2C1\%2C\%22Measure\%22\%2C\%22Measure\%22\%2C\%22Measure\%22\%2C\%22Measure\%22\%5D\%2C\%5B1\%2C1\%2C1\%2C\%22Z\%5E\%C2\%BC\%22\%2C\%22Z\%5E\%C2\%BC\%22\%2C\%22Z\%5E\%C2\%BC\%22\%2C\%22Z\%5E\%C2\%BC\%22\%2C\%22Z\%5E\%C2\%BC\%22\%2C\%22Z\%5E\%C2\%BC\%22\%2C\%22Z\%5E\%C2\%BC\%22\%2C\%22Z\%5E\%C2\%BC\%22\%5D\%2C\%5B1\%2C1\%2C1\%2C\%22H\%22\%2C\%22H\%22\%2C\%22H\%22\%2C\%22H\%22\%2C\%22H\%22\%2C\%22H\%22\%2C\%22H\%22\%2C\%22H\%22\%5D\%2C\%5B1\%2C1\%2C1\%2C\%22Measure\%22\%2C\%22Measure\%22\%2C\%22Measure\%22\%2C\%22Measure\%22\%2C\%22Measure\%22\%2C\%22Measure\%22\%2C\%22Measure\%22\%2C\%22Measure\%22\%5D\%2C\%5B\%22Z\%22\%2C\%22Z\%22\%2C\%22Z\%22\%2C\%22\%E2\%80\%A2\%22\%2C1\%2C1\%2C1\%2C1\%2C1\%2C1\%2C1\%2C1\%2C\%22X\%22\%5D\%2C\%5B\%22Z\%22\%2C\%22Z\%22\%2C1\%2C1\%2C\%22\%E2\%80\%A2\%22\%2C1\%2C1\%2C1\%2C1\%2C1\%2C1\%2C1\%2C\%22X\%22\%5D\%2C\%5B\%22Z\%22\%2C1\%2C\%22Z\%22\%2C1\%2C1\%2C\%22\%E2\%80\%A2\%22\%2C1\%2C1\%2C1\%2C1\%2C1\%2C1\%2C\%22X\%22\%5D\%2C\%5B\%22Z\%22\%2C1\%2C1\%2C1\%2C1\%2C1\%2C\%22\%E2\%80\%A2\%22\%2C1\%2C1\%2C1\%2C1\%2C1\%2C\%22X\%22\%5D\%2C\%5B1\%2C\%22Z\%22\%2C\%22Z\%22\%2C1\%2C1\%2C1\%2C1\%2C\%22\%E2\%80\%A2\%22\%2C1\%2C1\%2C1\%2C1\%2C\%22X\%22\%5D\%2C\%5B1\%2C\%22Z\%22\%2C1\%2C1\%2C1\%2C1\%2C1\%2C1\%2C\%22\%E2\%80\%A2\%22\%2C1\%2C1\%2C1\%2C\%22X\%22\%5D\%2C\%5B1\%2C1\%2C\%22Z\%22\%2C1\%2C1\%2C1\%2C1\%2C1\%2C1\%2C\%22\%E2\%80\%A2\%22\%2C1\%2C1\%2C\%22X\%22\%5D\%2C\%5B1\%2C1\%2C1\%2C1\%2C1\%2C1\%2C1\%2C1\%2C1\%2C1\%2C\%22\%E2\%80\%A2\%22\%2C1\%2C\%22X\%22\%5D\%2C\%5B\%22Z\%22\%2C1\%2C1\%2C1\%2C1\%2C1\%2C1\%2C1\%2C1\%2C1\%2C1\%2C\%22\%E2\%80\%A2\%22\%5D\%2C\%5B1\%2C1\%2C\%22Z\%22\%2C1\%2C1\%2C1\%2C1\%2C1\%2C1\%2C1\%2C1\%2C1\%2C1\%2C\%22\%E2\%80\%A2\%22\%5D\%2C\%5B1\%2C\%22Z\%22\%2C1\%2C1\%2C1\%2C1\%2C1\%2C1\%2C1\%2C1\%2C1\%2C1\%2C1\%2C1\%2C\%22\%E2\%80\%A2\%22\%5D\%2C\%5B\%22X\%22\%2C\%22X\%22\%2C\%22X\%22\%5D\%2C\%5B\%22Amps3\%22\%2C1\%2C1\%2C1\%2C1\%2C1\%2C1\%2C1\%2C1\%2C1\%2C1\%2C1\%2C\%22\%7C0\%E2\%9F\%A9\%E2\%9F\%A80\%7C\%22\%5D\%5D\%7D}{following this link}.
    Discerning readers can follow the link and edit the circuit in order to confirm that adding a single Z error by any T gate is caught by the post-selection, and also that all possible pairs of Z errors escape detection.
  }
\end{figure*}

\begin{figure*}
  \label{fig:ccz-slices-simple}
  \resizebox{\textwidth}{!}{
    \includegraphics{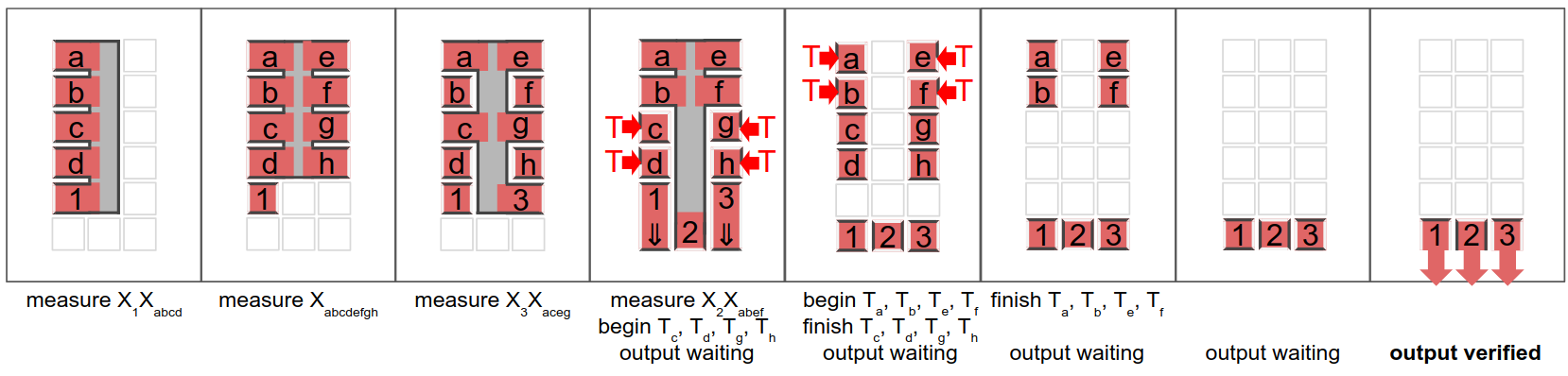}
  }
  \caption{
    Time slices of lattice surgery activity during production of a single $|CCZ\rangle$ state.
    Each red square corresponds to a qubit, and the label inside the red square identifies the qubit from \fig{ccz-circuit} that the square corresponds to.
    Gray rectangles correspond to X stabilizer measurements between sets of qubits.
    The red arrows labelled ``T" correspond to a noisy T state entering the system.
    It is possible to double the throughput shown here by interleaving the production of two states (shown in \fig{ccz-slices}).
  }

  \resizebox{\textwidth}{!}{
    \includegraphics{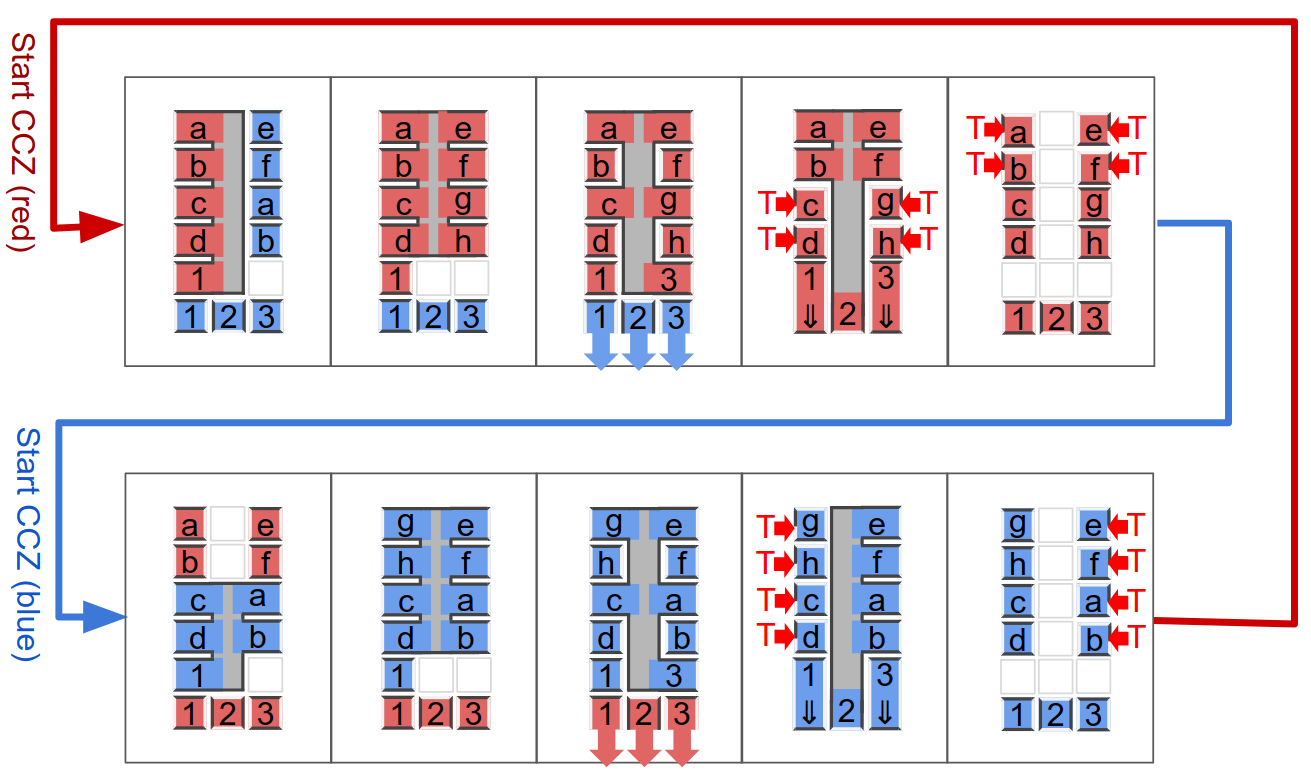}
  }
  \caption{
    Time slices of lattice surgery activity during production of $|CCZ\rangle$ states by the \factory{8|T\rangle}{28\epsilon^2}{|CCZ\rangle}.
    To maximize utilization, two states are produced concurrently.
    Each red or blue square corresponds to a qubit, and the label inside the square identifies the qubit from \fig{ccz-circuit} that the square corresponds to.
    Gray rectangles correspond to X stabilizer measurements between sets of qubits.
    The red arrows labelled ``T" correspond to a noisy T state entering the system.
    Blue squares correspond to qubits involved in producing one of the states, and red squares correspond to qubits involved in producing the other state.
    The red squares in each step are exactly identical to the red squares shown in the matching step of \fig{ccz-slices-simple}.
    See \fig{ccz-3d} for a 3D topological diagram corresponding to the time slices.
  }
  \label{fig:ccz-slices}
\end{figure*}

\begin{figure*}
  \label{fig:ccz-3d}
  \includegraphics[width=\textwidth,height=\dimexpr\textheight-14\baselineskip,keepaspectratio]{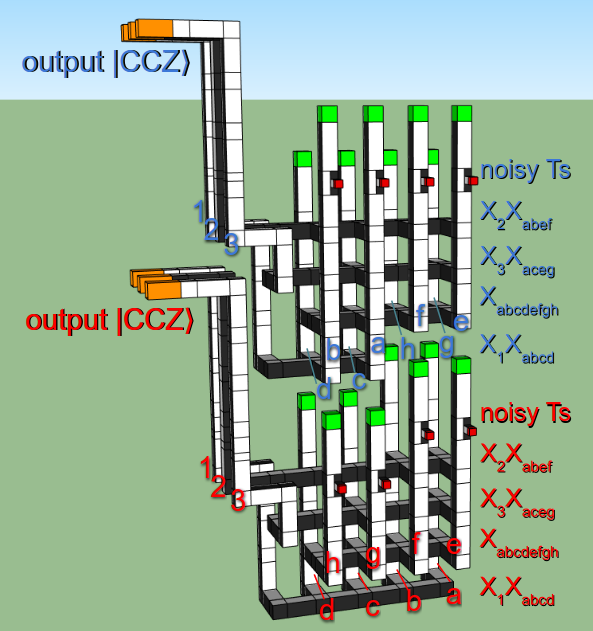}
  \caption{
    3D topological diagram for our construction of the \factory{8|T\rangle}{28\epsilon^2}{|CCZ\rangle}.
    The spacing between qubits has been increased to make it possible to see the internal structure.
    White (black) surfaces correspond to boundaries where chains of Z (X) errors can terminate.
    Corresponds to the time slices from \fig{ccz-slices}.
    Time increases from bottom to top.
    The vertical poles correspond to qubits from \fig{ccz-circuit}.
    Red boxes indicate connection points for noisy $|T_1\rangle$ states produced by a level 1 T factory.
    The green boxes atop the columns are performing either an X or Y basis measurement at half code distance, as described in \cite{fowler2018}, by including or omitting an S gate performed using twists \cite{brown2017poking}.
    Using half code distance is acceptable because, at the location in the factory where these operations are performed (i.e. after the T injections), individual errors are detected as distillation failures.
    The labels along the right hand side indicate the stabilizer measurements occurring at each time.
    The red/blue coloring of labels matches the red/blue coloring of \fig{ccz-slices}.
    Note that inserting the $|T_1\rangle$ state has a depth of 1.5, unlike the other steps which have depth 1.
    Each horizontal bar linking several vertical poles is a stabilizer measurement of a product of logical X observables.
    The groups of three qubits highlighted orange and exiting left are the $|CCZ\rangle$ states being output (note that the middle pole of each $|CCZ\rangle$ state is rotated with respect to the others, with white on top instead of black on top).
    The two instances of the factory that are shown differ slightly.
    Their qubits have been permuted so that each factory's top layer fits into a void at the bottom of the following factory, saving a layer of depth.
  }
\end{figure*}

\begin{figure*}
  \label{fig:ccz-graph}
  \resizebox{\textwidth}{!}{
    \includegraphics{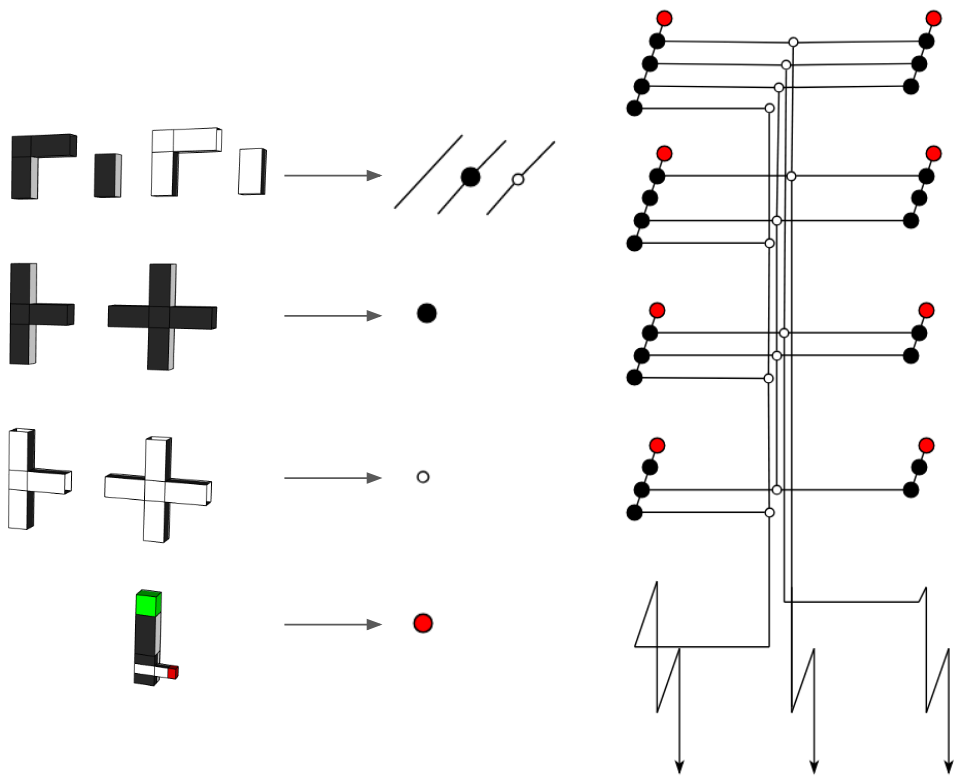}
  }
  \caption{
    A substitution procedure (left) for translating our 3D topological diagrams into (nearly) the ZX calculus \cite{de2017}, as well as an example translation of one of the CCZ factories from \fig{ccz-3d} (right).
    We use black (white) nodes instead of green (red) nodes (the usual notation for the ZX calculus) so that the node colors match the boundary colors in the 3D topological diagrams.
    Pieces with two ports are translated into edges or degree-2 nodes of either color.
    Pieces with three or more ports are translated into a node of matching color.
    The $Z \otimes Z$ measurement of a qubit vs a $|T\rangle$ state followed by measuring the qubit in the X or Y basis depending on the outcome of the parity measurement is translated into a (non-standard) red node.
    The red node can be expanded into a proper ZX calculus construction, but we do not attempt to do so.
    The ZX calculus graph is more amenable to verification than the 3D diagram.
    The reverse translation, from ZX calculus graph to 3D topological diagram, is often more difficult.
  }
\end{figure*}

A key technique introduced in \cite{fowler2018} is a single-layer stabilizer measurement involving an arbitrary numbers of qubits.
We use this technique in order to quickly measure the 4 stabilizers of the error-detecting Toffoli distillation protocol \cite{jones2013, eastin2013distilling}.
See \fig{ccz-circuit} for a circuit diagram of the CCZ-distillation process.
The operations in the circuit are chosen in a way that trivially translates into lattice surgery.
In \fig{ccz-slices-simple} we show time slices of one possible translation of the circuit into lattice surgery (with matching qubit labels and operation labels), and then in \fig{ccz-slices} show the time slices of our CCZ factory (corresponding to two interleaved translations of the circuit).
We also provide an annotated 3D topological diagram of the CCZ factory (see \fig{ccz-3d}).

Our $|CCZ\rangle$ factory has a naive depth of $4$ (stabilizer measurements) + $1.5$ (T state injections) + $1$ (X or Y basis measurement, depending on T injection measurements) + $2$ (detect errors) = $8.5$.
We use the same technique as in \cite{fowler2018} to partially overlap executions of the factory, resulting in an effective depth of $5.5$.
The T state injections take $1.5d$ layers because they are performed at half code distance and it takes $0.5d$ layers to move the black side into position for a parity measurement, then $0.5d$ layers to perform the parity measurement, then $0.5d$ layers to return the black side to its original position.
It is acceptable to inject at half code distance because the incoming T states have an error rate larger than the topological error incurred from an injection at this distance.

Our $|CCZ\rangle$ factory produces magic states fast enough that algorithms will bottleneck on routing instead of magic state production unless special care is taken.
For example, suppose there are several Toffoli operations to perform on qubits all placed in a common area; a common area with exactly one entrance capable of allowing exactly one qubit to enter or leave every $d$ cycles.
Because a new $|CCZ\rangle$ state is produced every $5.5d$ cycles, and each such state involves three qubits, the entrance will be occupied for $3d$ out of every $5.5d$ cycles moving magic state qubits into the common area to meet target qubits (or vice versa). This leaves only $2.5d$ cycles for other work requiring the entrance.
Furthermore, the $|CCZ\rangle$ teleportation process requires classically controlled CZ and CNOT operations.
If these operations also block the entrance, and are not done in a way that minimizes depth, they will use up the remaining $2.5d$ cycles and cause a routing bottleneck.

We see three ways for algorithms to avoid bottlenecking on routing and keep up with our $|CCZ\rangle$ factory:

\begin{enumerate}
\item
    Increase the amount of space dedicated to routing.
    Play it safe; do not have areas with narrow entrances or hallways that can only accommodate one qubit per $d$ cycles.
    This strategy is simple and effective, but costly.
\item
    Carefully distribute logical qubits across multiple disjoint areas with the goal of ensuring that Toffolis rarely target multiple qubits in the same area.
    This avoids the bottleneck by having the magic state qubits pass through multiple different entrances, instead of one common entrance.
    This strategy will not work for all algorithms, but it will work for some algorithms.
\item
    Use generalized CCZ operations capable of targeting arbitrary stabilizers instead of individual qubits, and move Clifford work into the classical control system.
    The generalized CCZ is performed in the same way that \cite{litinski2018} performs generalized T gates targeting arbitrary stabilizers.
    The gate teleportation process is modified; replacing each $Z_t \otimes Z_m$ parity measurement between a target qubit $t$ and the magic state qubit $m$ with a many-body stabilizer measurement $P \otimes Z_m$ where $P$ is a vector of Pauli operations possibly involving every logical data qubit in the computation.
    The main drawback of this approach is that there is 2x space overhead associated with ensuring it is always fast to access the $X$, $Y$, and $Z$ observable of every qubit.
    This can likely be avoided by interleaving single-qubit work between the Toffoli operations, but requires careful algorithm-by-algorithm consideration.
\end{enumerate}

Note that our CCZ factory's footprint includes an unused 2x4 area, adjacent to where the $|CCZ\rangle$ state exits the factory (see \fig{overview-dataflow}).
This area can be used to hold target qubits waiting for a Toffoli operation, which helps with the routing overhead. Our overhead spreadsheet assumes this space will be used in this manner.

In order to produce a $|CCZ\rangle$ state every $5.5d$ cycles, we need enough level 1 T factories to create 8 $|T\rangle$ states every $5.5d$ cycles.
The half-code-distance level 1 T factory from \cite{fowler2018} produces a $|T\rangle$ state every $3.25d$ cycles, except when distillation errors are detected.
Assuming a physical gate error rate of $10^{-3}$ and a level 1 code distance of 15, distillation errors will be detected approximately 3\% of the time (the $|T_0\rangle$ states have $\sim 10^{-3}$ error when injected, gain $\sim 10^{-3}$ error while the level 0 T gates are performed at distance 7, there are fifteen of them, and the most likely case is that a single one fails: $2 \cdot 10^{-3} \cdot 15 = 3\%$).
These failures reduce the effective output rate to a $|T\rangle$ state every $3.35d$ cycles, so five of these factories will produce $\sim 8.2$ $|T\rangle$ states every $5.5d$ cycles, which is sufficient to keep up with the $|CCZ\rangle$ factory.
We accumulate a buffer of surplus level 1 $|T\rangle$ states in the small hallways between the $|CCZ\rangle$ factory and the level 1 $|T\rangle$ factories so that a single level 1 T factory failure does not delay the entire $|CCZ\rangle$ factory.
As shown in \fig{overview-dataflow}, the five level 1 factories are placed to either side of the $|CCZ\rangle$ factory.
Note that it is occasionally necessary to route the fifth factory's output to the opposite side, and that there is enough contiguous unused volume in the factory to do this when needed.

We compute the error rate of the $|CCZ\rangle$ states being produced by our factory in two different regimes: the large code distance regime where the factory is distillation limited, and the minimal code distance regime where the factory may be limited by topological errors in the surface code.
We assume a physical gate error rate of $10^{-3}$ in both cases, and assume that the post-selected state injection of Li \cite{li2015} creates $|T_0\rangle$ states with approximately this probability of error.
In the distillation limited regime, we run these states through the \factory{15|T\rangle}{35 \epsilon^3}{|T\rangle} and then through our \factory{8|T\rangle}{28\epsilon^2}{|CCZ\rangle} producing intermediate $|T_1\rangle$ states with error rate $\sim 3.5 \cdot 10^{-8}$ and then $|CCZ\rangle$ states with error rate $\sim 3.4 \cdot 10^{-14}$.
In the minimal code distance regime, we must account for topological error introduced while performing T gates and the Clifford operations making up the factory.
For example, we assume that the error rate of the $|T_0\rangle$ states doubles while performing a level 0 T gate at distance 7.
This increases the effective error of the $|T_1\rangle$ states, but this contribution is overshadowed by the large size and proportionally small code distance of the level 1 T factory operating on these states.
The factory adds approximately $10^{-6}$ error to the output error, which is three to four times more than the distillation error.
We sum the two error rates, resulting in an estimated error rate for $|T_1\rangle$ states of $\sim 1.4 \cdot 10^{-6}$.
This is forty times more error than in the distillation limited case.
The CCZ factory has a code distance large enough that we are distillation limited, and the error rate of the final $|CCZ\rangle$ states is correspondingly $\sim 5.3 \cdot 10^{-11}$.

As shown in figure \fig{spreadsheet}, the minimal distance factory causes errors in more than 50\% of runs when attempting to factor a 1024 bit number, but can comfortably run classically intractable chemistry algorithms.
However, if one increases the level 1 code distance from 15 to 19 (increasing the footprint of the factory by roughly 20\%), then the level 1 error improves so much that it's possible to factor 4096 bit numbers.

\section{\texorpdfstring{
    The $|T\rangle$-catalyzed $|CCZ\rangle \rightarrow 2|T\rangle$ factory
}{
    The T-catalyzed CCZ to 2T Factory
}}
\label{sec:catalysis}

\begin{figure*}
    \label{fig:catalysis-circuit-simple}
    \centering
    \resizebox{\linewidth}{!}{
        \includegraphics{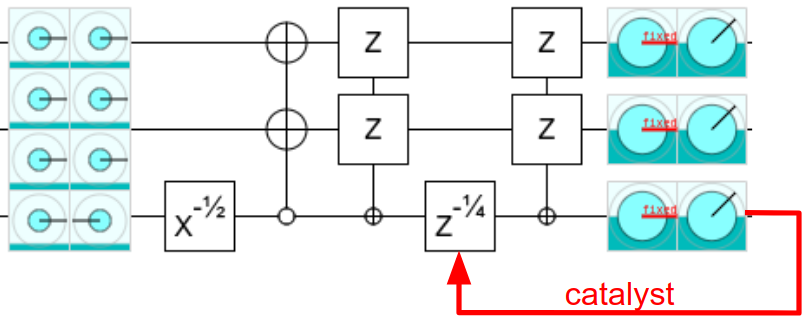}
    }
    \caption{
      A circuit that transforms a $|CCZ\rangle$ state into three $|T\rangle$ states by applying Clifford operations and a single T gate.
      By using one of the outputs to fuel the next iteration, the circuit can be re-interpreted as a circuit that turns one $|CCZ\rangle$ into two $|T\rangle$ states when catalyzed by one $|T\rangle$ state.
      The boxes with blue circles are state displays from the online simulator Quirk, with each circle representing an amplitude (the radius of the colored circle indicates the amplitude's magnitude, and the angle of the line rooted at the center of the circle indicates the phase).
      The state displays are showing that the input state is a $|CCZ\rangle$ and the output states are $|T\rangle$ states.
      The small circled pluses are X-axis controls (equivalent to a normal control surrounded by Hadamard gates).
      The circuit can be opened in Quirk by \href{http://algassert.com/quirk\#circuit=\%7B\%22cols\%22\%3A\%5B\%5B\%22H\%22\%2C\%22H\%22\%2C\%22H\%22\%5D\%2C\%5B\%22\%E2\%80\%A2\%22\%2C\%22\%E2\%80\%A2\%22\%2C\%22Z\%22\%5D\%2C\%5B\%22Amps3\%22\%5D\%2C\%5B\%5D\%2C\%5B1\%2C1\%2C\%22X\%5E-\%C2\%BD\%22\%5D\%2C\%5B\%22X\%22\%2C\%22X\%22\%2C\%22\%E2\%97\%A6\%22\%5D\%2C\%5B\%22Z\%22\%2C\%22Z\%22\%2C\%22\%E2\%8A\%96\%22\%5D\%2C\%5B1\%2C1\%2C\%22Z\%5E-\%C2\%BC\%22\%5D\%2C\%5B\%22Z\%22\%2C\%22Z\%22\%2C\%22\%E2\%8A\%96\%22\%5D\%2C\%5B\%22Amps1\%22\%2C\%22Amps1\%22\%2C\%22Amps1\%22\%5D\%5D\%7D}{following this link}.
    }
\end{figure*}

\begin{figure*}
    \label{fig:catalysis-circuit}
    \centering
    \includegraphics[width=\textwidth,height=\dimexpr\textheight-11\baselineskip,keepaspectratio]{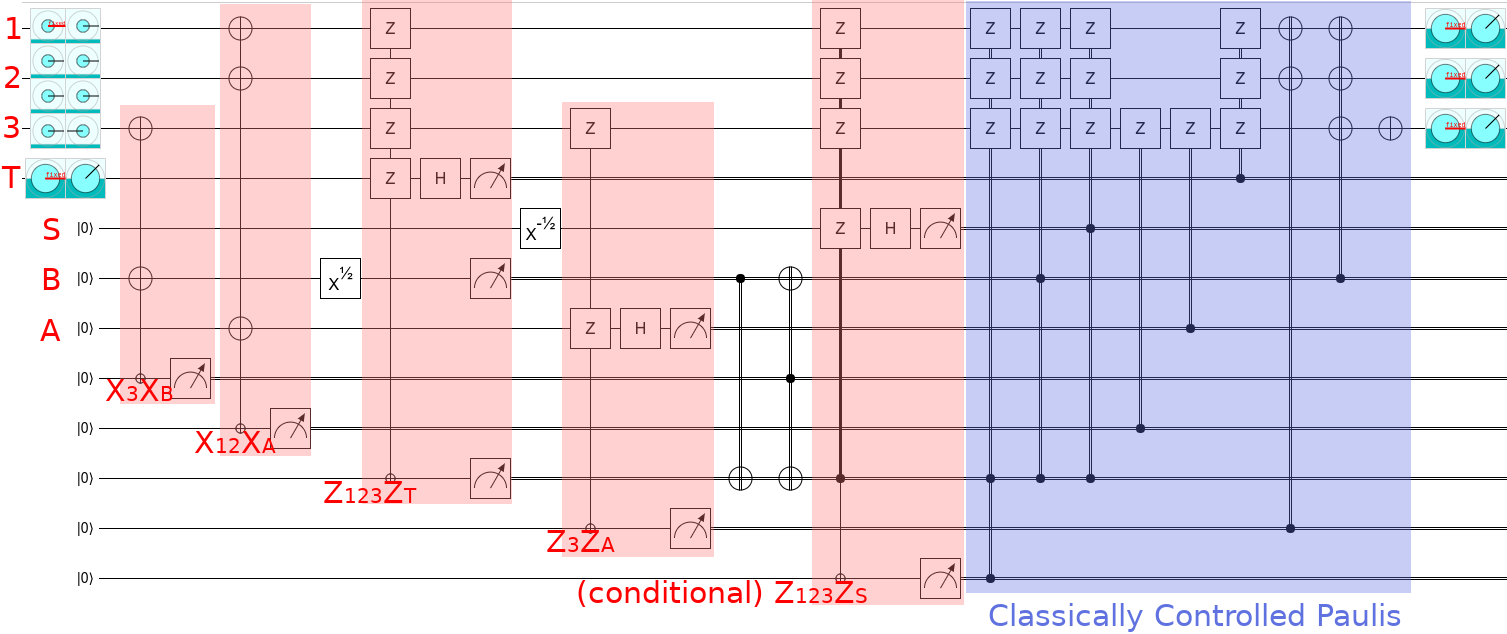}
    \caption{
      A circuit for catalyzed $|T\rangle$ state production, specialized for lattice surgery.
      Given a $|CCZ\rangle$ state (first three qubits) and a $|T\rangle$ state (fourth qubit), produces three $|T\rangle$ states.
      Red areas correspond to a product-of-Paulis measurement.
      The blue area happens entirely within classical control software.
      The $S$ ancilla is preparing an $|S\rangle$ state that can be used to correct the T gate teleportation used to perform the $Z^{-1/4}$ gate from \fig{catalysis-circuit-simple}.
      The $B$ ancilla is being used to perform the $X^{-1/2}$ gate from \fig{catalysis-circuit-simple}.
      The $A$ ancilla is being used to perform the multi-target CNOT from \fig{catalysis-circuit-simple}.
      The boxes with blue circles, at the beginning and end of the circuit, are state displays from the online simulator Quirk.
      Each circle represents an amplitude (the radius of the colored circle indicates the amplitude's magnitude, and the angle of the line rooted at the center of the circle indicates the phase).
      The state displays are showing that the input and output states are $|CCZ\rangle$ and $|T\rangle$ states as described.
      The small circled pluses in the circuit are X-axis controls (equivalent to a normal control surrounded by Hadamard gates); whenever one of these controls directly precedes a measurement the measurement corresponds to a Pauli product measurement.
      The circuit can be opened in Quirk by \href{http://algassert.com/quirk\#circuit=\%7B\%22cols\%22\%3A\%5B\%5B\%22H\%22\%2C\%22H\%22\%2C\%22H\%22\%2C\%22H\%22\%5D\%2C\%5B1\%2C1\%2C1\%2C\%22Z\%5E\%C2\%BC\%22\%5D\%2C\%5B\%22\%E2\%80\%A2\%22\%2C\%22\%E2\%80\%A2\%22\%2C\%22Z\%22\%5D\%2C\%5B\%22Amps3\%22\%2C1\%2C1\%2C\%22Amps1\%22\%5D\%2C\%5B\%5D\%2C\%5B1\%2C1\%2C\%22X\%22\%2C1\%2C1\%2C\%22X\%22\%2C1\%2C\%22\%E2\%8A\%96\%22\%5D\%2C\%5B1\%2C1\%2C1\%2C1\%2C1\%2C1\%2C1\%2C\%22Measure\%22\%5D\%2C\%5B\%22X\%22\%2C\%22X\%22\%2C1\%2C1\%2C1\%2C1\%2C\%22X\%22\%2C1\%2C\%22\%E2\%8A\%96\%22\%5D\%2C\%5B1\%2C1\%2C1\%2C1\%2C1\%2C1\%2C1\%2C1\%2C\%22Measure\%22\%5D\%2C\%5B1\%2C1\%2C1\%2C1\%2C1\%2C\%22X\%5E\%C2\%BD\%22\%5D\%2C\%5B\%22Z\%22\%2C\%22Z\%22\%2C\%22Z\%22\%2C\%22Z\%22\%2C1\%2C1\%2C1\%2C1\%2C1\%2C\%22\%E2\%8A\%96\%22\%5D\%2C\%5B1\%2C1\%2C1\%2C\%22H\%22\%5D\%2C\%5B1\%2C1\%2C1\%2C\%22Measure\%22\%2C1\%2C\%22Measure\%22\%2C1\%2C1\%2C1\%2C\%22Measure\%22\%5D\%2C\%5B1\%2C1\%2C1\%2C1\%2C\%22X\%5E-\%C2\%BD\%22\%5D\%2C\%5B1\%2C1\%2C\%22Z\%22\%2C1\%2C1\%2C1\%2C\%22Z\%22\%2C1\%2C1\%2C1\%2C\%22\%E2\%8A\%96\%22\%5D\%2C\%5B1\%2C1\%2C1\%2C1\%2C1\%2C1\%2C\%22H\%22\%5D\%2C\%5B1\%2C1\%2C1\%2C1\%2C1\%2C1\%2C\%22Measure\%22\%2C1\%2C1\%2C1\%2C\%22Measure\%22\%5D\%2C\%5B1\%2C1\%2C1\%2C1\%2C1\%2C\%22\%E2\%80\%A2\%22\%2C1\%2C1\%2C1\%2C\%22X\%22\%5D\%2C\%5B1\%2C1\%2C1\%2C1\%2C1\%2C\%22X\%22\%2C1\%2C\%22\%E2\%80\%A2\%22\%2C1\%2C\%22X\%22\%5D\%2C\%5B\%22Z\%22\%2C\%22Z\%22\%2C\%22Z\%22\%2C1\%2C\%22Z\%22\%2C1\%2C1\%2C1\%2C1\%2C\%22\%E2\%80\%A2\%22\%2C1\%2C\%22\%E2\%8A\%96\%22\%5D\%2C\%5B1\%2C1\%2C1\%2C1\%2C\%22H\%22\%5D\%2C\%5B1\%2C1\%2C1\%2C1\%2C\%22Measure\%22\%2C1\%2C1\%2C1\%2C1\%2C1\%2C1\%2C\%22Measure\%22\%5D\%2C\%5B\%22Z\%22\%2C\%22Z\%22\%2C\%22Z\%22\%2C1\%2C1\%2C1\%2C1\%2C1\%2C1\%2C\%22\%E2\%80\%A2\%22\%2C1\%2C\%22\%E2\%80\%A2\%22\%5D\%2C\%5B\%22Z\%22\%2C\%22Z\%22\%2C\%22Z\%22\%2C1\%2C1\%2C\%22\%E2\%80\%A2\%22\%2C1\%2C1\%2C1\%2C\%22\%E2\%80\%A2\%22\%5D\%2C\%5B\%22Z\%22\%2C\%22Z\%22\%2C\%22Z\%22\%2C1\%2C\%22\%E2\%80\%A2\%22\%2C1\%2C1\%2C1\%2C1\%2C\%22\%E2\%80\%A2\%22\%5D\%2C\%5B1\%2C1\%2C\%22Z\%22\%2C1\%2C1\%2C1\%2C1\%2C1\%2C\%22\%E2\%80\%A2\%22\%5D\%2C\%5B1\%2C1\%2C\%22Z\%22\%2C1\%2C1\%2C1\%2C\%22\%E2\%80\%A2\%22\%5D\%2C\%5B\%22Z\%22\%2C\%22Z\%22\%2C\%22Z\%22\%2C\%22\%E2\%80\%A2\%22\%5D\%2C\%5B\%22X\%22\%2C\%22X\%22\%2C1\%2C1\%2C1\%2C1\%2C1\%2C1\%2C1\%2C1\%2C\%22\%E2\%80\%A2\%22\%5D\%2C\%5B\%22X\%22\%2C\%22X\%22\%2C\%22X\%22\%2C1\%2C1\%2C\%22\%E2\%80\%A2\%22\%5D\%2C\%5B1\%2C1\%2C\%22X\%22\%5D\%2C\%5B\%22Amps1\%22\%2C\%22Amps1\%22\%2C\%22Amps1\%22\%5D\%5D\%7D}{following this link}.
    }
\end{figure*}

\begin{figure*}
    \label{fig:catalysis-slices}
    \centering
    \resizebox{\linewidth}{!}{
        \includegraphics{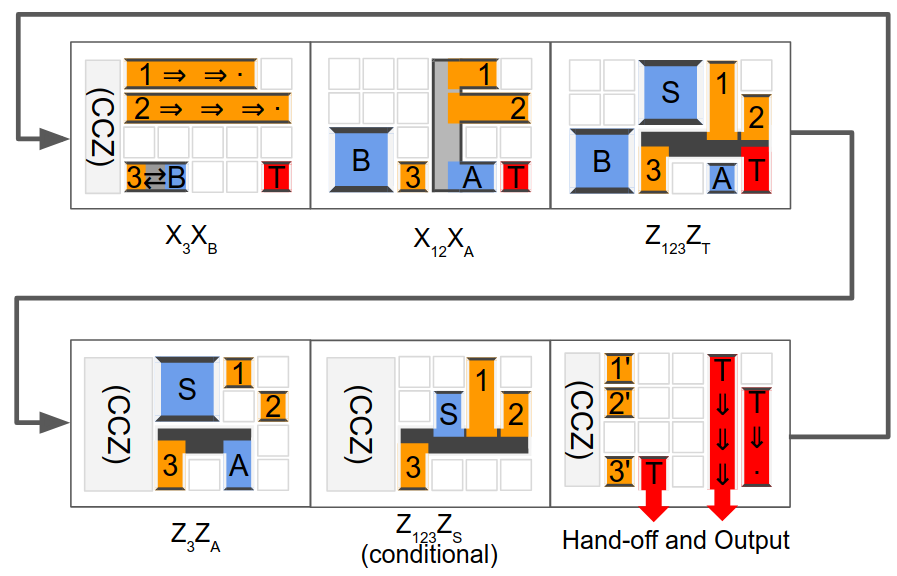}
    }
    \caption{
        Time slices of lattice surgery activity during transformation of a $|CCZ\rangle$ state (orange qubits labelled 1, 2, 3) into three $|T\rangle$ states (shown in red in last slice), catalyzed by a $|T\rangle$ state (bottom right qubit in red).
        Black and dark gray bars correspond to stabilizer measurements.
        Ancillae qubits are shown in blue.
        The code distance of the ancillae qubits is doubled when single-qubit Clifford operations are being applied, to ensure there is sufficient suppression of errors.
        The light gray ``(CCZ)" box to the left will be used by the CCZ factory producing $|CCZ\rangle$ states to be transformed.
        See \fig{catalysis-3d} for a 3D topological diagram corresponding to the time slices.
        Every step being performed can be matched up with a step from \fig{catalysis-circuit}, and the qubit labels shown here correspond to the qubit labels there.
    }
\end{figure*}

\begin{figure*}
  \label{fig:catalysis-3d}
  \includegraphics[width=\textwidth,height=\dimexpr\textheight-6\baselineskip,keepaspectratio]{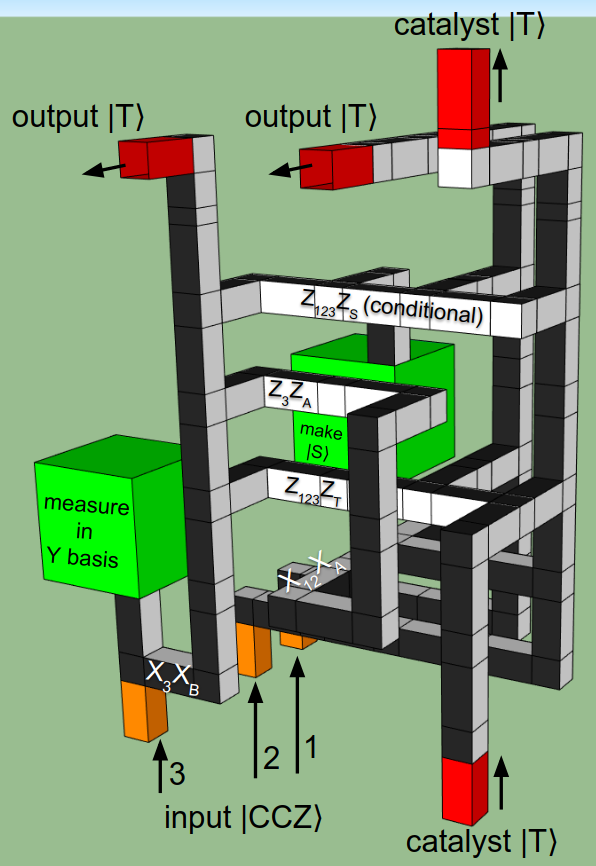}
  \caption{
    3D topological diagram of a lattice surgery circuit transforming a $|CCZ\rangle$ state (orange-tipped inputs at bottom) and a $|T\rangle$ state (bottom right red-tipped input) into three $|T\rangle$ states (red-tipped outputs at top).
    We conservatively assume that the green boxes are large enough to perform any single-qubit Clifford with negligible error.
    See \fig{ccz-3d} for details about how to interpret the diagram.
  }
\end{figure*}

In \cite{jones2013}, it is shown how to perform a Toffoli gate by using Clifford operations, measurement, and four T gates.
That circuit can be rewritten into an inline circuit that transforms three $|+\rangle$ states into a $|CCZ\rangle$ state via Clifford operations and four T gates \cite{gidney2018}.
Then, by diagonalizing that circuit's stabilizer table, said circuit can be rewritten into a form where three of the T gates apply directly to an input $|+\rangle$ state.
Those three T gates can then be replaced by three $|T\rangle$ state inputs, resulting in a circuit that maps $|T\rangle^{\otimes 3}$ to $|CCZ\rangle$ using Clifford gates and one T gate.
This circuit contains no measurement, and therefore can be inverted.
The inverse circuit (shown in \fig{catalysis-circuit-simple}) maps a $|CCZ\rangle$ state to three $|T\rangle$ states using Clifford gates and one T gate.

Because $|T\rangle$ states can be used to perform T gates, the T gate used to transform the $|CCZ\rangle$ into three $|T\rangle$ states can be powered by a $|T\rangle$ state output from a previous iteration of the circuit.
If we keep feeding a $|T\rangle$ state output from iteration $k$ into iteration $k+1$, then we effectively have a circuit that takes a $|CCZ\rangle$ state and outputs two $|T\rangle$ states.
Under this interpretation of the circuit, the third $|T\rangle$ state is an ancillary state that is necessary for the transformation to be possible, but is not consumed by the transformation.
Thus, in keeping with terminology for ancillary states that enable LOCC communication tasks without being consumed \cite{jonathan1999}, and previous work \cite{campbell2011catalysis}, we refer to the third $|T\rangle$ as a catalyst.
We refer to the circuit as a whole as the $|T\rangle$-catalyzed $|CCZ\rangle \rightarrow 2|T\rangle$ factory, or ``C2T factory" for short.

Beware that, although the catalyst $|T\rangle$ state is not consumed by the C2T factory, it does accumulate noise from the incoming $|CCZ\rangle$ states.
If a catalyst $|T\rangle$ has cycled through $n$ iterations of the C2T factory, and there is a probability $\epsilon$ of each $|CCZ\rangle$ containing an error, then there is an $\Theta(n \epsilon)$ chance that the catalyst has been poisoned and is causing the factory to produce bad outputs.
However, because every error in the catalyst ultimately traces back to an error in a $|CCZ\rangle$ state, the chance of there being {\em any} error grows like $\Theta(n \epsilon)$, instead of $\Theta(n^2 \epsilon)$ as would be expected from a naive calculation assuming uncorrelated errors.

Distillation protocols usually require inputs with uncorrelated errors, so it is important that we only use the C2T factory as the last step in a distillation chain.
In a sense, because of how we use the C2T factory, the correlation between errors is beneficial to us instead of detrimental.
It means that when we run an algorithm many times there will be a small number of runs with many errors, instead of many runs with a small number of errors.
We experience quadratically fewer whole-algorithm failures than would be expected from the fact that the expected number of errors is growing like $\Theta(n^2 \epsilon)$.
For other examples of correlation between errors being beneficial, we recommend reviewing hat guessing games \cite{paterson2010}.

The C2T factory circuit shown in \fig{catalysis-circuit-simple} is compact, but not in an ideal form for embedding into lattice surgery.
\fig{catalysis-circuit} fixes this by providing an equivalent circuit that, although it appears much more complicated, trivially translates into lattice surgery.
We show the result of this translation in \fig{catalysis-slices}, which has time slices of the lattice surgery operations occurring as the factory operates.
And finally \fig{catalysis-3d} shows an annotated 3D topological diagram of the process.

\section{Arbitrary-Angle Phase Catalysis}
\label{sec:generalize}

\begin{figure*}
    \label{fig:catalysis-circuit-generalized}
    \centering
    \resizebox{\linewidth}{!}{
        \includegraphics{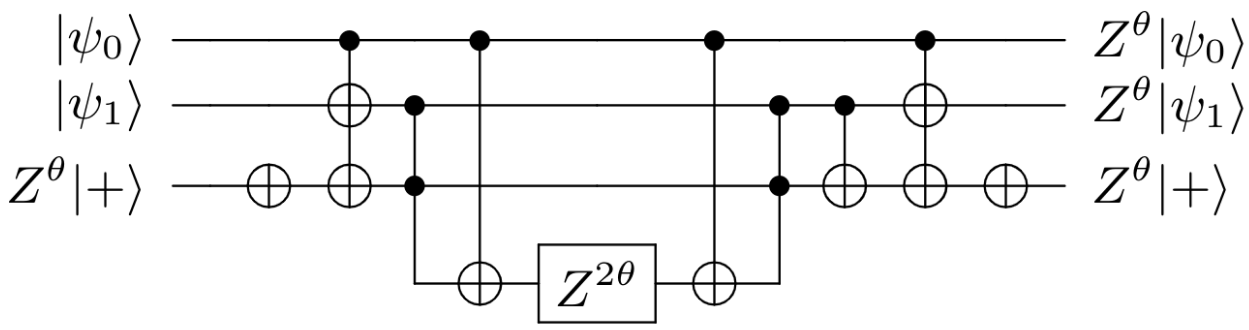}
    }
    \caption{
        Generalized phase catalysis circuit.
        Given a $Z^\theta |+\rangle$ catalyst, two $Z^\theta$ operations can be applied via stabilizer gates, one AND computation gate (notation from \cite{gidney2018}), and one $Z^{2 \theta}$ gate.
    }
\end{figure*}

The catalysis technique used in the C2T factory from the previous section generalizes to phasing angles other than the T gate's $45^\circ$.
In \fig{catalysis-circuit-generalized}, we show a generalization of \fig{catalysis-circuit-simple} that works for an arbitrary angle $\theta$.
This circuit performs two $Z^\theta$ operations by performing cheap stabilizer operations, performing one Toffoli gate, performing one $Z^{2 \theta}$ operation, and being catalyzed by one $Z^\theta |+\rangle$ state.
Contrast with gate teleportation \cite{gottesman1999}, which consumes a previously prepared $Z^\theta |+\rangle$ state in order to perform one $Z^\theta$ operation, with a 50\% chance of requiring a fixup $Z^{-2 \theta}$ operation.

One way to discover the generalized phase catalysis circuit is to start from the phase-gradient-via-addition circuit \cite{kitaev2002, gidney2018, nam2018}, which performs a series of rotations $Z$, $S$, $T$, $\sqrt{T}$, $\sqrt{\sqrt{T}}$, etc by adding a register containing the target qubits into a phase gradient catalyst state.
Include a carry bit input in the addition of the phase-gradient-via-addition circuit, truncate the circuit after the first ripple-carry step by using the correct fixup operation, and the result is a phase catalysis circuit for an angle $\theta=\pi/2^k$ which trivially generalizes to arbitrary angles.
The catalysis circuit can likely also be derived from synthillation parity-check circuits \cite{campbell2018}, which use similar magic states and have a similar structure but are used to perform distillation of existing states instead of producing additional states.

\begin{figure*}
    \label{fig:catalysis-circuit-sqrt-t}
    \centering
    \resizebox{\linewidth}{!}{
        \includegraphics{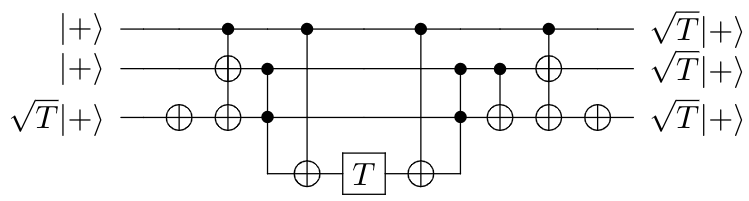}
    }
    \caption{
        Using a catalyst $|\sqrt{T}\rangle$ state to create 2 additional $|\sqrt{T}\rangle$ states using cheap stabilizer operations, one T gate, and one AND computation.
        Has a T-cost of $5$ \cite{jones2013, gidney2018}, implying the T-cost of a $\sqrt{T}$ state is at most $2.5$.
    }
\end{figure*}

Specializing the generalized phase catalysis circuit to $\theta = 22.5^{\circ}$, i.e. to the $\sqrt{T}$ gate, produces the circuit shown in \fig{catalysis-circuit-sqrt-t}.
This specialized circuit creates two $|\sqrt{T}\rangle$ states by performing one Toffoli operation and one T gate.
This is significantly more efficient than previous techniques we were able to find and adapt to the task of producing $|\sqrt{T}\rangle$ states \cite{landahl2013complex, bocharov2014, mishra2014, kitaev2002, gidney2018, nam2018}, assuming a physical gate error rate of $10^{-3}$ and a target error rate of $10^{-10}$.
For example, according to figure 5 of \cite{bocharov2014}, repeat-until-success circuits use $\approx 45$ T gates to approximate a $\sqrt{T}$ gate to within precision $\epsilon = 10^{-10}$.
As another example, according to table III of \cite{mishra2014}, direct synthesis of $\sqrt{T}$ state uses $\approx 25$ times more volume than direct synthesis of $|T\rangle$ states (though this ratio improves as the physical gate error rate improves).
A final example: the phase-gradient-via-addition operation described in \cite{kitaev2002,gidney2018} can perform a $\sqrt{T}$ gate with a 4-bit adder (which requires 3 $|CCZ\rangle$ states).
Phase-gradient-via-addition is the closest to competing with phase catalysis, which is perhaps not surprising since phase catalysis is an optimized form of this technique.
Other techniques appear to be very far behind; requiring an order of magnitude more spacetime volume.

\section{\texorpdfstring{
Lattice surgery construction of the \factory{8|T\rangle}{28\epsilon^2}{2|T\rangle}
}{
Lattice surgery construction of the 8T to 2T factory
}}
\label{sec:full}

We now combine the $|CCZ\rangle$ factory from \sec{ccz} with the C2T factory from \sec{catalysis}, producing a $|T\rangle$-catalyzed T factory that transforms eight noisy $|T\rangle$ states into two $|T\rangle$ states with quadratically less noise.
Note that this means we achieve a 4:1 ratio of input $|T\rangle$ states to output $|T\rangle$ state, which is competitive with the 3:1 ratio of block codes \cite{bravyi2012}.
This is surprising, because normally one has to work with a larger number of $|T\rangle$ states in order to achieve good ratios.

Note that we do not use exactly the same CCZ factory as in \sec{ccz}.
We re-order the stabilizer measurements and place the output qubits in a different location, so that it fits into the C2T factory from \sec{catalysis}.
Furthermore, we do not bother interleaving the factory with itself anymore.
There's no point; we need five $T_1$ factories to run at the rate achieved by interleaving but now only have four factories (recall \fig{overview-dataflow}).

The details of the combined factory are covered in \fig{full-slices}, which shows the parallel operation of C2T factory and CCZ factory.
Recall that the qubit labels can be matched up with \fig{ccz-circuit} for verification that the correct stabilizers are being measured (though in a different order).
Our penultimate figure, \fig{full-3d}, shows a 3D topological diagram of the factory.
Note that the figure omits the level 1 T factories feeding in noisy $|T_1\rangle$ states, and exaggerates the spacing between qubits in order to make internal structures visible, but is otherwise complete.

To bootstrap the factory, an initial catalyst $|T\rangle$ state is made ``the hard way", using some less efficient $|T\rangle$ factory that can output $|T\rangle$ states with error no higher than the error rate of the $|CCZ\rangle$ factory.
Bootstrapping occurs once at the start of the computation, and any time the catalyst $|T\rangle$ state is lost.
Specifically, note that the $|CCZ\rangle$ state produced by the CCZ part of the factory is being consumed before it's known if it contained a distillation error.
Therefore, when a detected distillation error does occur, the $|T\rangle$ state catalyst must be discarded.
This has a negligible effect on the effective depth of the factory, because it occurs so rarely (approximately once per hundred thousand distillations).
There is a space towards the top right of the factory where a spare $|T\rangle$ catalyst could be placed, to be used as a backup when the main catalyst is lost.

The primary bottleneck on the output of this factory is the rate at which $|T_1\rangle$ states are produced.
As shown in \fig{overview-dataflow}, we assume there are four $|T_1\rangle$ factories present (one beside each pair of qubits require $|T_1\rangle$ states.
When functioning perfectly, each of these factories produces a pair of noisy $|T_1\rangle$ states every $6.5d$ cycles, which is just enough to feed the catalyzed T factory and keep it producing a pair of $|T_2\rangle$ states every $6.5d$ cycles.

\begin{figure*}
    \label{fig:t1-3d}
    \centering
    \includegraphics[width=\textwidth,height=\dimexpr\textheight-12\baselineskip,keepaspectratio]{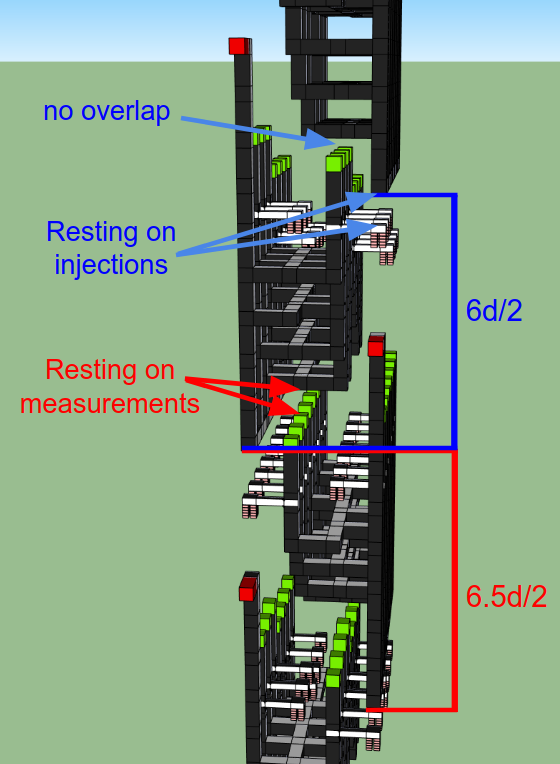}
    \caption{
        3D topological diagram of a re-arrangement of the level 1 T factory from \cite{fowler2018}.
        Quantities are quoted in units of $d/2$ instead of $d$ because the factory is performed at half code distance.
        Improves the depth from $13d/2$ to $12.5d/2$, increasing the output rate by roughly 4\%, which ensures four level 1 T factories is sufficient to feed our T-catalyzed factory.
        There are two variants of the factory: the one shown with an output on the left (A) and the one shown with an output on the right (B).
        The qubits of A and B have been permuted so that their first stabilizer measurement involves qubits that are all on the same side, allowing the stabilizer measurement to be performed without using a central bar.
        The second measurement of B (the first measurement using the central bar) is over the back 8 qubits and the last measurement of A is over the front 8 qubits.
        This allows B to be lowered by half of $d/2$, so that B rests on the level 0 $|T\rangle$ injections to the right of A.
        The transition back from B to A cannot be lowered quite as far, because the top of B would intersect the first central bar used in A.
        Overall this optimization saves $0.5d/2$ depth relative to the interleaving technique used in \cite{fowler2018}.
    }
\end{figure*}

\begin{figure*}
  \label{fig:full-slices}
  \resizebox{\textwidth}{!}{
    \includegraphics{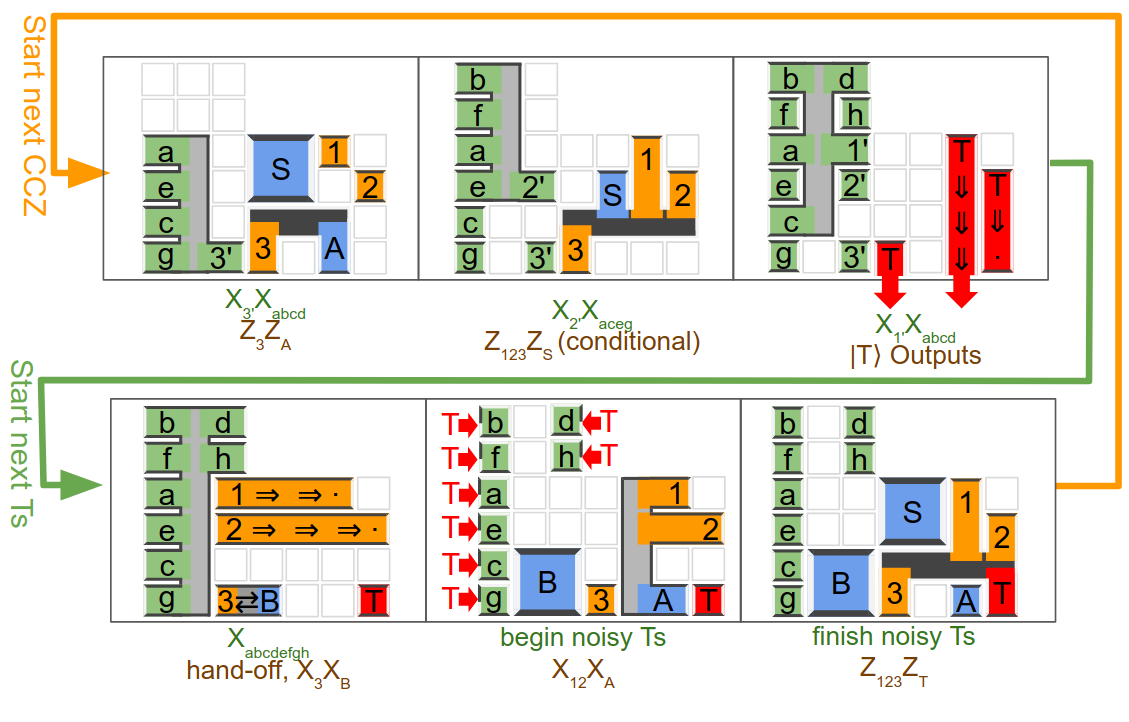}
  }
  \caption{
    Time slices of activity during catalyzed $|T\rangle$ state distillation.
    Every step being performed can be matched up with \fig{ccz-circuit} and \fig{catalysis-slices}.
    See \fig{full-3d} for the 3D topological diagram these time slices come from.
  }
\end{figure*}

\begin{figure*}
  \label{fig:full-3d}
    \includegraphics[width=\textwidth,height=\dimexpr\textheight-6\baselineskip,keepaspectratio]{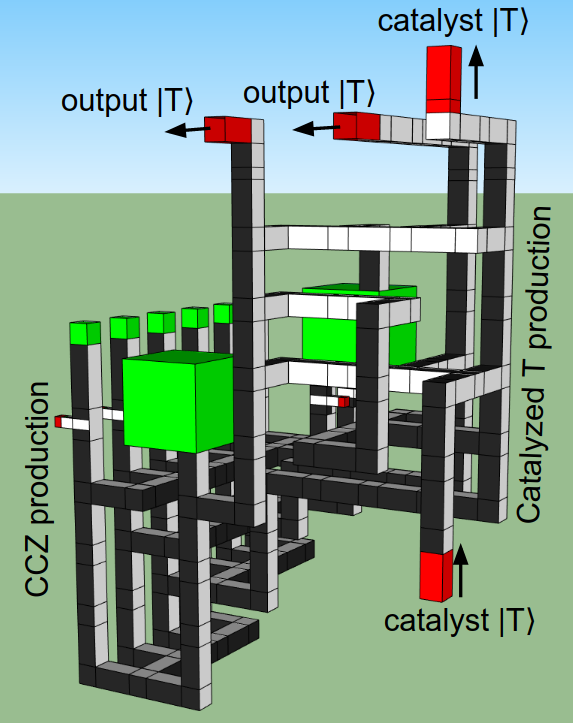}
  \caption{
    3D topological diagram of the full $|T\rangle$-catalyzed \factory{8|T\rangle}{28\epsilon^2}{2|T\rangle}.
    Contrast with the time slices from \fig{full-slices}, and the circuit in \fig{ccz-circuit} combined with the circuit in \fig{catalysis-circuit-simple}.
    Single-qubit Clifford gates that would affect the catalyst $|T\rangle$ if they failed are performed with extremely conservative code distance (large green boxes).
  }
\end{figure*}

Of course, the $|T_1\rangle$ factories do not always function perfectly.
They discard their output roughly 3\% of the time due to detecting an error (computed in \sec{ccz}).
In order to actually achieve a depth of $6.5d$ for the \factory{8|T\rangle}{28\epsilon^2}{2|T\rangle}, it is necessary increase the $|T_1\rangle$ factory output rate by more than 3\% to compensate.
There are many ways to achieve such a small gain, and \fig{t1-3d} sketches one way to do so.
Therefore the $T_1$ factories can keep up with the catalyzed T factory producing a pair of $|T_2\rangle$ states every $6.5d$ cycles.

\section{Conclusions}
\label{sec:conclusion}

\begin{figure*}
  \label{fig:continuous-operation-3d}
    \includegraphics[width=\textwidth,height=\dimexpr\textheight-6\baselineskip,keepaspectratio]{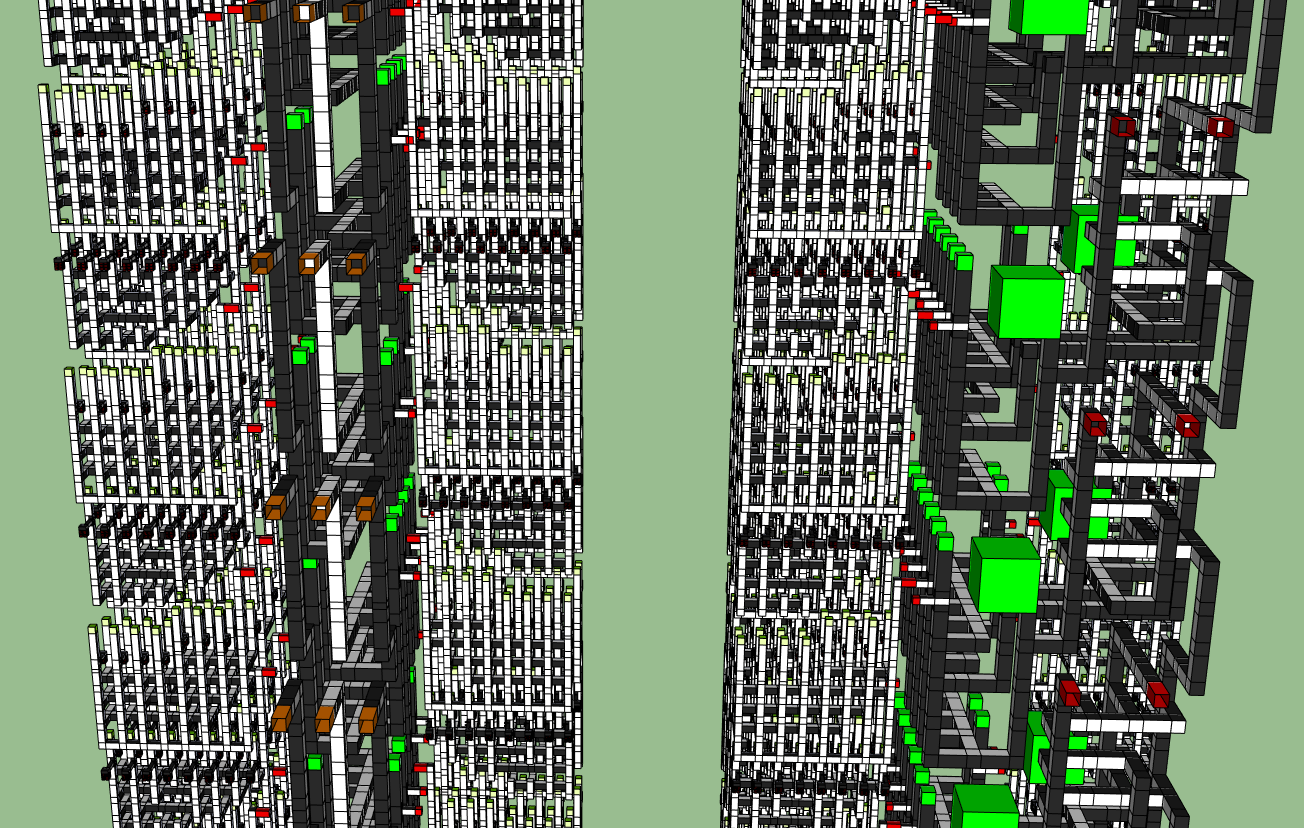}
  \caption{
    3D topological diagram showing tiled operation of our CCZ factory (left) and our catalyzed T factory (right).
    Includes the half-distance level 1 T factories feeding noisy states into the larger factories.
  }
\end{figure*}

In this paper we presented two factories: a $|CCZ\rangle$ factory and a catalyzed $|T\rangle$ factory.
We compiled these factories all the way down to 3d topological diagrams (see \fig{continuous-operation-3d}) and gave detailed estimates of their spacetime volume, footprint, and error rates.
We also showed how to generalize the phase catalysis technique used by our $|T\rangle$ factory to apply to arbitrary angles, including the particularly-efficient angle of $\theta=22.5^\circ$.
Finally, we slightly improved the output rate of the level 1 T factories from \cite{fowler2018}, gave a simple procedure for transforming topological diagrams into ZX calculus graphs, provided a resource estimator spreadsheet, and gave working simulator links for verifying most of our circuit constructions.

Because it takes four $|T\rangle$ states to perform a Toffoli gate, but only one $|CCZ\rangle$ state to do the same, algorithms dominated by applying Toffolis, such as Shor's algorithm and the chemistry algorithm in \cite{babbush2018}, run five times as fast when using our $|CCZ\rangle$ factory instead of the $|T\rangle$ factory from \cite{fowler2018}.
However, we caution that it is often necessary to rework these algorithms' circuits to account for the much faster Toffoli rate.
Assuming that such a reworking is possible for \cite{babbush2018}, the runtimes at classically intractable sizes would be reduced from $\sim$10 hours (see table VII of \cite{babbush2018}) to $\sim$2 hours.
For algorithms dominated by performing T gates, our catalyzed T factory provides a more modest 2$\times$ speedup.

We believe it is possible to further decrease the volume of our factories.
For example, we suspect that the level 1 T state injection at the end of each factory can be partially merged with that factory's final stabilizer measurement.
If that is true, then the depth of the factories could be reduced by $1d$.
However, the effect of this optimization on the topological error rate is difficult to predict and we will use simulation to check the optimization's correctness before claiming it.

Another possible optimization is to eagerly route $|CCZ\rangle$ qubits emerging from the CCZ factory to their final destination (in preparation for a parity measurement), instead of holding them next to the factory until they are verified.
Removing the output-holding area reduces the CCZ factory's footprint by over 20\%, which is a large gain, but it is important to keep in mind that this is not a true reduction in volume but rather a reclassification of some of the factory's volume as routing volume.

Yet another possible optimization would be to carefully analyze how topological errors within the surface code propagate through the factory.
At any location where an error chain between two boundaries would result in a detected failure, the boundaries can be moved closer together.

A final idea that should be investigated is estimating logical error probabilities from the observed pattern of detection events produced by the surface code's stabilizer measurements.
For example, if there were a sudden burst of detection events crossing between two boundaries during the execution of a factory, the factory's output could be cautiously discarded even if the logical measurement results indicate there is no problem.
Assuming there is some metric that can be derived from the raw detection events, that reliably correlates with the true failure probability, this would allow us to reduce the number of false negatives (where an undetected error escapes the factory) at the cost increasing the number of false positives (where a run with no error is discarded).

In this paper we focused on making a low-volume factory in the single-factory regime, but it is also important to consider factories optimized to have a tiny footprint.
Early quantum computers will have limited space; it's worth sacrificing depth if it means the factory actually fits on the machine.
By combining techniques from this paper and low-footprint distillation techniques mentioned in \cite{litinski2018}, it should be possible to create factories covering fewer qubits but with roughly the same volume as ours.

Another interesting avenue to explore is the high-footprint / multi-factory regime, where factories based on block codes become possible.
Block factories should be able to outperform the efficiency of our factory, assuming enough states are being distilled in parallel.
But this raises the question of whether block factories can also be improved by catalysis; are there catalyzed block factories?
We don't know the answer to this question.

We suspect that the space of quantum circuits contains many other gems akin to the catalyzed phasing circuit.
We consider finding these circuit to be important, because they can be surprisingly efficient at their tasks.
It would be particularly useful to have a general framework for finding catalyzed circuits, to better understand what makes them efficient, and to understand the connection with related constructions such as distillation via parity-checks \cite{campbell2018}, synthillation \cite{campbell2017}, and phase gradient kickbacks \cite{kitaev2002, gidney2018, nam2018}.
Our guess as to the nature of the connection between these constructions is that there are a small number of circuit identities underlying all these related but different techniques, and that each technique is rewriting and interpreting the underlying circuit identities in a different way.
If the connection is actually of this form, then perhaps it is possible to write code that takes a circuit using one of these techniques, derives the identity the circuit is using, and then produces a whole related family of interesting circuits (perhaps including circuits that use catalysis).

\section{Acknowledgements}

We thank Earl T. Campbell for reviewing a draft of this paper and suggesting useful references.

\bibliographystyle{plainnat}
\bibliography{refs}

\begin{thebibliography}{32}
\providecommand{\natexlab}[1]{#1}
\providecommand{\url}[1]{\texttt{#1}}
\expandafter\ifx\csname urlstyle\endcsname\relax
  \providecommand{\doi}[1]{doi: #1}\else
  \providecommand{\doi}{doi: \begingroup \urlstyle{rm}\Url}\fi

\bibitem[Babbush et~al.(2018)Babbush, Gidney, Berry, Wiebe, McClean, Paler,
  Fowler, and Neven]{babbush2018}
Ryan Babbush, Craig Gidney, Dominic~W Berry, Nathan Wiebe, Jarrod McClean,
  Alexandru Paler, Austin Fowler, and Hartmut Neven.
\newblock Encoding electronic spectra in quantum circuits with linear t
  complexity.
\newblock \emph{arXiv preprint arXiv:1805.03662}, 2018.
\newblock \doi{10.1103/PhysRevX.8.041015}.

\bibitem[Bocharov et~al.(2015)Bocharov, Roetteler, and Svore]{bocharov2014}
Alex Bocharov, Martin Roetteler, and Krysta~M Svore.
\newblock Efficient synthesis of universal repeat-until-success quantum
  circuits.
\newblock \emph{Physical review letters}, 114\penalty0 (8):\penalty0 080502,
  2015.
\newblock \doi{10.1103/PhysRevLett.114.080502}.

\bibitem[Bravyi and Kitaev(1998)]{Brav98}
S.~B. Bravyi and A.~Yu. Kitaev.
\newblock Quantum codes on a lattice with boundary.
\newblock \emph{quant-ph/9811052}, 1998.
\newblock URL \url{https://arxiv.org/abs/quant-ph/9811052}.

\bibitem[Bravyi and Haah(2012)]{bravyi2012}
Sergey Bravyi and Jeongwan Haah.
\newblock Magic-state distillation with low overhead.
\newblock \emph{Physical Review A}, 86\penalty0 (5):\penalty0 052329, 2012.
\newblock \doi{10.1103/PhysRevA.86.052329}.

\bibitem[Bravyi and Kitaev(2005)]{bravyi2005}
Sergey Bravyi and Alexei Kitaev.
\newblock Universal quantum computation with ideal clifford gates and noisy
  ancillas.
\newblock \emph{Physical Review A}, 71\penalty0 (2):\penalty0 022316, 2005.
\newblock \doi{10.1103/PhysRevA.71.022316}.

\bibitem[Brown et~al.(2017)Brown, Laubscher, Kesselring, and
  Wootton]{brown2017poking}
Benjamin~J Brown, Katharina Laubscher, Markus~S Kesselring, and James~R
  Wootton.
\newblock Poking holes and cutting corners to achieve clifford gates with the
  surface code.
\newblock \emph{Physical Review X}, 7\penalty0 (2):\penalty0 021029, 2017.
\newblock \doi{10.1103/PhysRevX.7.021029}.

\bibitem[Campbell(2011)]{campbell2011catalysis}
Earl~T Campbell.
\newblock Catalysis and activation of magic states in fault-tolerant
  architectures.
\newblock \emph{Physical Review A}, 83\penalty0 (3):\penalty0 032317, 2011.
\newblock \doi{10.1103/PhysRevA.83.032317}.

\bibitem[Campbell and Howard(2017)]{campbell2017}
Earl~T Campbell and Mark Howard.
\newblock Unified framework for magic state distillation and multiqubit gate
  synthesis with reduced resource cost.
\newblock \emph{Physical Review A}, 95\penalty0 (2):\penalty0 022316, 2017.
\newblock \doi{10.1103/PhysRevA.95.022316}.

\bibitem[Campbell and Howard(2018)]{campbell2018}
Earl~T Campbell and Mark Howard.
\newblock Magic state parity-checker with pre-distilled components.
\newblock \emph{Quantum}, 2:\penalty0 56, 2018.
\newblock \doi{10.22331/q-2018-03-14-56}.

\bibitem[de~Beaudrap and Horsman(2017)]{de2017}
Niel de~Beaudrap and Dominic Horsman.
\newblock The zx calculus is a language for surface code lattice surgery.
\newblock \emph{arXiv preprint arXiv:1704.08670}, 2017.
\newblock URL \url{https://arxiv.org/abs/1704.08670}.

\bibitem[Dennis et~al.(2002)Dennis, Kitaev, Landahl, and Preskill]{Denn02}
E.~Dennis, A.~Kitaev, A.~Landahl, and J.~Preskill.
\newblock Topological quantum memory.
\newblock \emph{J. Math. Phys.}, 43:\penalty0 4452--4505, 2002.
\newblock \doi{10.1063/1.1499754}.
\newblock quant-ph/0110143.

\bibitem[Eastin(2013)]{eastin2013distilling}
Bryan Eastin.
\newblock Distilling one-qubit magic states into toffoli states.
\newblock \emph{Physical Review A}, 87\penalty0 (3):\penalty0 032321, 2013.
\newblock \doi{10.1103/PhysRevA.87.032321}.

\bibitem[Fowler et~al.(2012)Fowler, Mariantoni, Martinis, and Cleland]{Fowl12f}
A.~G. Fowler, M.~Mariantoni, J.~M. Martinis, and A.~N. Cleland.
\newblock Surface codes: Towards practical large-scale quantum computation.
\newblock \emph{Phys. Rev. A}, 86:\penalty0 032324, 2012.
\newblock \doi{10.1103/physreva.86.032324}.
\newblock arXiv:1208.0928.

\bibitem[Fowler and Devitt(2012)]{fowler2012bridge}
Austin~G Fowler and Simon~J Devitt.
\newblock A bridge to lower overhead quantum computation.
\newblock \emph{arXiv preprint arXiv:1209.0510}, 2012.
\newblock URL \url{https://arxiv.org/abs/1209.0510}.

\bibitem[Fowler and Gidney(2018)]{fowler2018}
Austin~G Fowler and Craig Gidney.
\newblock Low overhead quantum computation using lattice surgery.
\newblock \emph{arXiv preprint arXiv:1808.06709}, 2018.
\newblock URL \url{https://arxiv.org/abs/1808.06709}.

\bibitem[Fowler et~al.(2013)Fowler, Devitt, and Jones]{fowler2013}
Austin~G Fowler, Simon~J Devitt, and Cody Jones.
\newblock Surface code implementation of block code state distillation.
\newblock \emph{Scientific reports}, 3:\penalty0 1939, 2013.
\newblock \doi{10.1038/srep01939}.

\bibitem[Gidney(2018)]{gidney2018}
Craig Gidney.
\newblock Halving the cost of quantum addition.
\newblock \emph{Quantum}, 2:\penalty0 74, 2018.
\newblock \doi{10.22331/q-2018-06-18-74}.

\bibitem[Gottesman and Chuang(1999)]{gottesman1999}
Daniel Gottesman and Isaac~L Chuang.
\newblock Demonstrating the viability of universal quantum computation using
  teleportation and single-qubit operations.
\newblock \emph{Nature}, 402\penalty0 (6760):\penalty0 390, 1999.
\newblock \doi{10.1038/46503}.

\bibitem[Horsman et~al.(2012)Horsman, Fowler, Devitt, and
  Van~Meter]{horsman2012}
Clare Horsman, Austin~G Fowler, Simon Devitt, and Rodney Van~Meter.
\newblock Surface code quantum computing by lattice surgery.
\newblock \emph{New Journal of Physics}, 14\penalty0 (12):\penalty0 123011,
  2012.
\newblock \doi{10.1088/1367-2630/14/12/123011}.

\bibitem[Jonathan and Plenio(1999)]{jonathan1999}
Daniel Jonathan and Martin~B Plenio.
\newblock Entanglement-assisted local manipulation of pure quantum states.
\newblock \emph{Physical Review Letters}, 83\penalty0 (17):\penalty0 3566,
  1999.
\newblock \doi{10.1103/PhysRevLett.83.3566}.

\bibitem[Jones(2013)]{jones2013}
Cody Jones.
\newblock Low-overhead constructions for the fault-tolerant toffoli gate.
\newblock \emph{Physical Review A}, 87\penalty0 (2):\penalty0 022328, 2013.
\newblock \doi{10.1103/PhysRevA.87.022328}.

\bibitem[Kitaev et~al.(2002)Kitaev, Shen, Vyalyi, and Vyalyi]{kitaev2002}
Alexei~Yu Kitaev, Alexander Shen, Mikhail~N Vyalyi, and Mikhail~N Vyalyi.
\newblock \emph{Classical and quantum computation}.
\newblock American Mathematical Soc., 2002.

\bibitem[Landahl and Cesare(2013)]{landahl2013complex}
Andrew~J Landahl and Chris Cesare.
\newblock Complex instruction set computing architecture for performing
  accurate quantum $ z $ rotations with less magic.
\newblock \emph{arXiv preprint arXiv:1302.3240}, 2013.
\newblock URL \url{https://arxiv.org/abs/1302.3240}.

\bibitem[Li(2015)]{li2015}
Ying Li.
\newblock A magic state’s fidelity can be superior to the operations that
  created it.
\newblock \emph{New Journal of Physics}, 17\penalty0 (2):\penalty0 023037,
  2015.
\newblock \doi{10.1088/1367-2630/17/2/023037}.

\bibitem[Litinski(2019)]{litinski2018}
Daniel Litinski.
\newblock A game of surface codes: Large-scale quantum computing with lattice
  surgery.
\newblock \emph{Quantum}, 3:\penalty0 128, 2019.
\newblock \doi{10.22331/q-2019-03-05-128}.

\bibitem[Mishra and Fowler(2014)]{mishra2014}
Prashant Mishra and Austin Fowler.
\newblock Resource comparison of two surface code implementations of small
  angle z rotations.
\newblock \emph{arXiv preprint arXiv:1406.4948}, 2014.
\newblock URL \url{https://arxiv.org/abs/1406.4948}.

\bibitem[Nam et~al.(2018)Nam, Su, and Maslov]{nam2018}
Yunseong Nam, Yuan Su, and Dmitri Maslov.
\newblock Approximate quantum fourier transform with o(n log(n)) t gates.
\newblock \emph{arXiv preprint arXiv:1803.04933}, 2018.
\newblock URL \url{https://arxiv.org/abs/1803.04933}.

\bibitem[Paterson and Stinson(2010)]{paterson2010}
Maura~B Paterson and Douglas~R Stinson.
\newblock Yet another hat game.
\newblock \emph{the electronic journal of combinatorics}, 17\penalty0
  (1):\penalty0 R86, 2010.
\newblock URL
  \url{https://www.combinatorics.org/ojs/index.php/eljc/article/view/v17i1r86}.

\bibitem[Raussendorf and Harrington(2007)]{Raus07}
R.~Raussendorf and J.~Harrington.
\newblock Fault-tolerant quantum computation with high threshold in two
  dimensions.
\newblock \emph{Phys. Rev. Lett.}, 98:\penalty0 190504, 2007.
\newblock \doi{10.1103/PhysRevLett.98.190504}.
\newblock URL \url{https://doi.org/10.1103/PhysRevLett.98.190504}.
\newblock quant-ph/0610082.

\bibitem[Raussendorf et~al.(2007)Raussendorf, Harrington, and Goyal]{Raus07d}
R.~Raussendorf, J.~Harrington, and K.~Goyal.
\newblock Topological fault-tolerance in cluster state quantum computation.
\newblock \emph{New J. Phys.}, 9:\penalty0 199, 2007.
\newblock \doi{10.1088/1367-2630/9/6/199}.
\newblock URL \url{https://doi.org/10.1088/1367-2630/9/6/199}.
\newblock quant-ph/0703143.

\bibitem[Shor(1994)]{shor1994}
Peter~W Shor.
\newblock Algorithms for quantum computation: Discrete logarithms and
  factoring.
\newblock In \emph{Foundations of Computer Science, 1994 Proceedings., 35th
  Annual Symposium on}, pages 124--134. Ieee, 1994.
\newblock \doi{10.1109/SFCS.1994.365700}.

\bibitem[Zalka(1998)]{zalka1998fast}
Christof Zalka.
\newblock Fast versions of shor's quantum factoring algorithm.
\newblock \emph{arXiv preprint quant-ph/9806084}, 1998.
\newblock URL \url{https://arxiv.org/abs/quant-ph/9806084}.

\end{thebibliography}

\end{document}